 \documentclass[twocolumn]{aastex631}

\usepackage{graphicx}	
\usepackage{amsmath}	
\usepackage{booktabs}
\usepackage{xcolor}

\begin{document}

\title{Accretion properties of soft X-ray transient XTE J1856+053 during its 2023 Outburst}

\correspondingauthor{Debjit Chatterjee}
\email{debjitchatterjee92@gmail.com, debjit@mx.nthu.edu.tw}

\author[0000-0001-6770-8351]{Debjit Chatterjee}
\affiliation{Institute of Astronomy, National Tsing Hua University, Hsinchu 300044, Taiwan}
\affiliation{Indian Institute of Astrophysics, II Block Koramangala, Bangalore 560034, India}

\author{Arghajit Jana}
\affiliation{Instituto de Estudios Astrof{\'i}sicos, Universidad Diego Portales, Av. Ej{\'e}rcito Libertador 441, Santiago 8370191, Chile}
\affiliation{Institute of Astronomy, National Tsing Hua University, Hsinchu 300044, Taiwan}
\author{Hsiang-Kuang Chang}
\affiliation{Institute of Astronomy, National Tsing Hua University, Hsinchu 300044, Taiwan}
\affiliation{Department of Physics, National Tsing Hua University, Hsinchu 300044, Taiwan}

\begin{abstract}

Soft X-ray transients are a subclass of the low mass X-ray binaries that occasionally show a sudden rise in their soft X-ray luminosity; otherwise, they remain in an extremely faint state. We investigate the accretion properties of the soft X-ray transient XTE J1856+053 during its 2023 outburst obtained by {\it NICER} and {\it NuSTAR} data in July. We present detailed results on the timing and spectral analysis of the X-ray emission during the outburst. The power spectral density shows no quasi-periodic oscillation features. The source's spectrum on July 19 can be well-fitted with a multi-color blackbody component, a power-law component, and a reflection component with a broadened iron emission line. {\it NICER} spectra can be well-fitted by considering a combination of a blackbody and a power-law. The source exhibits a transition within just five days from a soft state to an intermediate state during the outburst decline phase. The inner accretion disk has a low inclination angle ($\sim18^\circ$). The spectral analysis also suggests a high-spin ($a>0.9$) BH as the central accreting object.

\end{abstract}

\keywords{accretion: accretion disks -- stars: individual: (XTE~J1856+053) -- X-rays: binaries.}


\section{Introduction} \label{sec:intro}
X-ray binaries (XRBs) are astrophysical systems emitting high levels of X-ray radiation, comprising a normal star and a compact object, which can be either a black hole (BH) or a neutron star (NS). The categorization of XRBs into high-mass (HMXBs) and low-mass (LMXBs) depends upon the mass of the companion star \citep{White1995,Remillard&McClintock2006}. Specifically, HMXBs feature companion stars of O- or B-type, while LMXBs involve A-type or later stars \citep{Tetarenko2016}. While HMXBs generally exhibit sustained brightness, a substantial portion of LMXBs display transient behavior. LMXBs may remain in a state of quiescence, falling below the detection threshold of monitoring instruments, for extensive periods before transitioning into a transient phase marked by a sudden increase of flux (called an outburst). This increase in flux endures for a span of weeks to months, followed by a gradual decline. An updated catalog \footnote{http://astro.uni-tuebingen.de/~xrbcat/}\footnote{https://binary-revolution.github.io/} of the LMXBs and HMXBs can be found in \cite{Fortin2022,Fortin2023,Avakyan2023,Neumann2023}.

During an outburst, the spectral composition of a black hole X-ray binary (BHXRB) is primarily attributed to two key components: a multi-color disk blackbody and a power-law tail. The thin Keplerian accretion disk \citep{SS73,Novikov&Thorne1973} around a BH is considered to be composed of annular regions with different temperatures, each radiating thermally, and its total spectrum is approximated by a multi-color disk blackbody \citep{Mitsuda1984}. The origin of the power law tail is considered to be from a hot Compton corona \citep{Sunyaev&Titarchuk1980,Sunyaev&Titarchuk1985}. Inverse Comptonization of the soft X-ray photons from the accretion disk by the hot electrons in the Compton cloud contributes to the high energy non-thermal emission from the XRB \citep{Haardt&Maraschi1993,Zdziarski1993,Titarchuk1994,CT95,Zycki1999}. As the outburst progresses, the relative contribution of these two primary components changes. These dynamic events encompass several distinctive spectral phases \citep{Fender2004,Remillard&McClintock2006,Done2007}.

At the onset of an outburst, the system enters a hard state (HS) characterized by a dominant hard power-law component, possibly accompanied by the presence of a steady jet \citep{Fender2004, Belloni2010}. As the rate of accretion intensifies, the source transitions through a hard and soft intermediate state (HIMS and SIMS) before eventually reaching a soft state (SS). In this latter phase, the accretion disk predominantly governs the emission, resulting in a considerably softer power-law component, while the jet activity subsides \citep{Miyamoto1991, Remillard2006}. This soft state persists until the late fading stages of the outburst, after which the system reverts to an intermediate state with a lower intensity compared to the preceding intermediate phase \citep{Homan2001}. Finally, at the end of the outburst, the system reverts back to the hard state. This cyclical progression traces a counterclockwise `q'-shaped trajectory in the hardness-intensity diagram (HID) \citep{Homan2005, Belloni2005, Nandi2012}.

Low-frequency quasi-periodic oscillations (LFQPOs) are a very common observable feature in the power density spectrum (PDS) of stellar-mass BHs. A few BHs exhibit high-frequency QPOs in their PDSs. The frequency of the QPOs in these transient X-ray sources ranges from mHz to a few hundred. In general, LFQPOs are classified into three types (A, B, C) based on the centroid frequency, Q-value, noise, and root-mean-square (rms) amplitude \citep{Casella2005}. It is observed that these LFQPOs correlated with the spectral states \citep{Remillard&McClintock2006}.

Not all BHXRB systems follow the typical `q' pattern observed in the HID during outbursts. Some transient systems undergo `hard-only' outbursts, where the source remains in the HS throughout the outburst or transitions only as far as the IMS without reaching the softer thermally dominant states (e.g.,\citealt{Hynes2000}, \citealt{Brocksopp2001}, \citealt{Belloni2002}, \citealt{Sidoli2011}, \citealt{Curran2013}, \citealt{DC2019}, \citealt{Alabarta2021}). This behavior is also observed in some persistently accreting systems, which either remain in the HS for extended periods or periodically undergo incomplete state transitions \citep{Churazov1993,delSanto2004,Soleri2012,Froning2014,Tetarenko2016,Debnath2020,DC2021b}.

Soft X-ray transients (SXTs) are a subclass of XRBs, characterized by their episodic and dramatic increases in soft (low-energy) X-ray luminosity \citep{Tanaka1996}. SXTs show a soft X-ray spectrum during outbursts, with the emission primarily from the accretion disk. Typically, SXTs consist of a compact object, such as a NS or a BH, accreting material from a companion star via an accretion disk \citep{Charles2006,Remillard2006}. SXTs, particularly those containing BHs, often remain undetected until an outburst occurs. During such an outburst, the soft X-ray luminosity of an SXT can increase by several orders of magnitude ($\sim10^{36}-10^{39}$~erg~s$^{-1}$), ranging from hundreds to thousands of times the quiescent level ($\sim10^{30}-10^{33}$~erg~s$^{-1}$) \citep{Chen1997,Esin1997,Lasota2001,Tetarenko2016}. Typically, these outbursts exhibit a rapid rise in luminosity over a few days, followed by a slower decline that spans approximately 30 days, often accompanied by a rebrightening phase in the X-ray light curve \citep{Chen1997}. These outbursts are believed to be triggered by thermal-viscous instabilities within the accretion disk, leading to rapid accretion onto the compact object and subsequent X-ray emission \citep{Lasota2001,Dubus2001}.

SXTs play a crucial role in understanding accretion processes, compact object formation, and the physics of extreme environments \citep{Remillard&McClintock2006}. The study of SXTs has provided insights into the nature of accretion disks, jet formation, and the properties of BHs and NSs \citep{Casares1992,McClintock2003,Fender2004}.

The SXT, XTE~J1856+053 was initially discovered during a survey of the Galactic ridge in 1996 using the {\it RXTE}/PCA instrument \citep{Marshall1996}. The source has exhibited intriguing outburst behavior over the years. The 1996 outburst displayed a distinctive light curve with two prominent peaks. The first peak, starting on April 4, 1996 (MJD 50177), demonstrated a symmetric rise and decline, lasting for 27 days and reaching a flux of 75 mCrab ($2.2\times10^{37}$~erg~s$^{-1}$) in the 2–12 keV energy range. Following this, the second peak, displaying a fast rise–slow decay (FRED) profile began on September 9, 1996 (MJD 50335), extended over 70 days, and reached its maximum flux of 79 mCrab ($2.3\times10^{37}$~erg~s$^{-1}$). This second X-ray peak was preceded by a precursor event 8 days earlier, featuring a flux of 30–60 mCrab ($0.9\times10^{37}-1.7\times10^{37}$~erg~s$^{-1}$) in higher energies (20–100 keV), as detected by {\it BATSE} on September 7–9, 1996 (MJD 50333–50335, \citealt{Barret1996}).

In February 2007, a new outburst of XTE~J1856+053 was observed, again with two peaks. This 2007 outburst was detected by {\it RXTE} \citep{Levine2007} and featured a precursor event occurring on January 10–15, 2007 (MJD 54110–54115). The first peak of the 2007 outburst began on February 28, 2007 (MJD 54159), achieving a maximum flux of approximately 85 mCrab on March 12, 2007 (MJD 54171), and extended for about 65 days. The subsequent peak started on May 21, 2007 (MJD 54241), exhibiting a rapid rise to approximately 110 mCrab within 7 days and lasting for approximately 55 days. Similar to the 1996 outburst, both peaks in 2007 were preceded by hard X-ray precursor events detected by {\it Swift}/BAT in the energy range of 10–200 keV, occurring from February 22 to March 1, 2007 (MJD 54153–54160), and on May 28–30, 2007 (MJD 54248–54250, \citealt{Krimm2007}). Furthermore, the hardness vs. intensity diagrams of both the 1996 and 2007 outbursts exhibited remarkable similarities.

A detailed study of the X-ray characteristics of XTE J1856+053 during its outburst events in 2007 using {\it XMM-Newton} Target of Opportunity (ToO) observation conducted on March 14, 2007, was done by \cite{Sala2008}. It is noted that the X-ray light curve for both outbursts exhibited two distinct peaks \citep{Sala2008}. 
The X-ray spectrum of XTE J1856+053 was modeled using a thermal accretion disk model, revealing a central temperature ($k{\rm T_{in}}$) of $0.75$ keV and a foreground absorption column density ($N_{\rm H}$) of $4.5\times10^{22}~{\rm cm}^{-2}$. The authors have considered infrared (IR) upper limits along with the high $N_{\rm H}$ value to infer that XTE~J1856+053 is most probably a BH.

Another outburst was reported in 2020 \citep{Negoro2015} using {\it MAXI}/GSC. The recent outburst was reported on July 12. 2023 by {\it MAXI}/GSC, and it is suggested that the outburst started on July 9, 2023 \citep{Kobayashi2023}.

In this paper, we study the outburst of XTE~J1856+053 in the 2023. The paper is organized in the following way. In the next section (\S2), we describe the observation and data reduction procedures. In Section \S3, we present the timing and spectral results. We discuss our results in Section \S4 and make a summary of the results in Section \S5.

\begin{table*}
	\caption{XTE J1856+053 observation log}
	\label{tab:table1}
	\begin{tabular}{lcccccr} 
		\hline
		Satellite/ & Obs ID & Date & MJD Start & MJD End & Avg MJD & Exposure\\
		Instrument &        & (dd-mm-yyyy) & & & &Time (ks)\\
        Col. 1 & Col. 2 & Col. 3 & Col. 4 & Col. 5 & Col. 6 & Col. 7\\
		\hline

		{\it NuSTAR} & 90901325002 (NuSTAR) & 19-07-2023 & 60144.065 &60144.771 &60144.418 &30.892\\
        {\it NICER} & 6100560101 (NICER1) &14-07-2023 & 60139.024& 60139.543& 60139.284 & 2.146\\
                    & 6100560102 (NICER2) &15-07-2023 & 60140.237& 60140.633& 60140.435 & 1.440\\
                    & 6100560103 (NICER3) &16-07-2023 & 60141.400& 60141.668& 60141.534 & 2.281\\
                    & 6100560104 (NICER4) &19-07-2023 & 60144.051& 60144.768& 60144.409 & 4.591\\
		\hline
	\end{tabular}
 
 \noindent{The details of the studied observations. The columns. 1, 2 \& 3 represent the satellites, observation IDs, and observation date, respectively. The start, the end and the average of the observations (in MJD) are given in Cols. 4, 5, and 6. The corresponding exposure time (in ks) is given in Col. 7. The only NuSTAR observation (90901325002) is combined with the NICER4 observation (6100560104) of the same day (July 19, 2023) for spectral study.}
\end{table*}

\section{Observations and Data Reduction} \label{sec:obs}
We used the archival data of four {\it NICER} observations and one {\it NuSTAR} observation of XTE~J1856+053 during its 2023 X-ray activity. The publicly available data were downloaded from the HeaSARC website \footnote{https://heasarc.gsfc.nasa.gov/cgi-bin/W3Browse/w3browse.pl}. The detailed observation log is given in Table~\ref{tab:table1}.

\subsection{NuSTAR}
{\it NuSTAR} observed XTE~J1856+053 on July 19, 2023. {\it NuSTAR} consists of two identical focal plane modules- FPMA and FPMB \citep{Harrison2013}. The {\it NuSTAR} raw data was reprocessed using the {\it NuSTAR Data Analysis Software} ({\tt NuSTARDAS version 2.1.2}). Calibrated and cleaned event files were generated by {\tt nupipeline} task. We used 20200912 version of calibration files from {\it NuSTAR} calibration database\footnote{http://heasarc.gsfc.nasa.gov/FTP/caldb/data/nustar/fpm/}.  We used 40'' circular regions for extracting the source, while the background region was selected with 60'' circle far from the source. The light curves and spectra were produced from the cleaned science mode event files through {\tt nuproducts} task. We rebinned the $3-78$ keV spectra with min 25 counts using {\tt grppha} task. Light curves were extracted with 100 sec time binning. The background-subtracted light curves from the two modules were combined with {\tt lcmath} task. 

\subsection{NICER}
After the reported outburst of XTE~J1856+053 on July 12, 2023, {\it NICER} observed the source on July 14, 15, 16, and 19. We used {\tt nicerl2}\footnote{https://heasarc.gsfc.nasa.gov/docs/nicer/analysis\_threads/nicerl2/} task to produce standard, calibrated, cleaned event files for each {\it NICER} observations. We used {\tt nicerl3-lc}\footnote{https://heasarc.gsfc.nasa.gov/docs/nicer/analysis\_threads/nicerl3-lc/} and {\tt nicerl3-spect}\footnote{https://heasarc.gsfc.nasa.gov/docs/nicer/analysis\_threads/nicerl3-spect/} for further extraction of the light curve and spectrum, respectively. During the generation of spectra, we used {\tt scorpeon} model for background calculation. The light curves were generated with 400~$\mu$s and 100~s binning. The spectra are grouped with a minimum of 25~counts/sec. PDSs were generated by applying the Fast Fourier Transformation (FFT) technique on the 400~$\mu s$ light curves using {\tt powspec} task of FTOOLS. We divided the light curves into 8192 intervals and computed the Poisson noise subtracted PDS for each interval. Then, we averaged all the PDS to obtain the final PDS for each observation. The final PDS are normalized to give the fractional rms spectra in $({\rm rms/mean})^2~{\rm Hz}^{-1}$ unit. Then, we geometric-binned the PDS with a factor of 0.07.

\section{Results}

We investigate the temporal and spectral properties of the soft X-ray transient XTE~J1856+053 during its 2023 X-ray outburst using {\it NICER} and {\it NuSTAR} data. Among the four {\it NICER} observations during this period, one observation (NICER4; Obs Id. 6100560104) coincides with the only NuSTAR observation (Obs Id. 90901325002) on July 19, 2023. Three more observations of {\it NICER} are taken from July 14 (NICER1), July 15 (NICER2), and July 16 (NICER3), 2023. 

\begin{figure}
	\includegraphics[width=8.5cm]{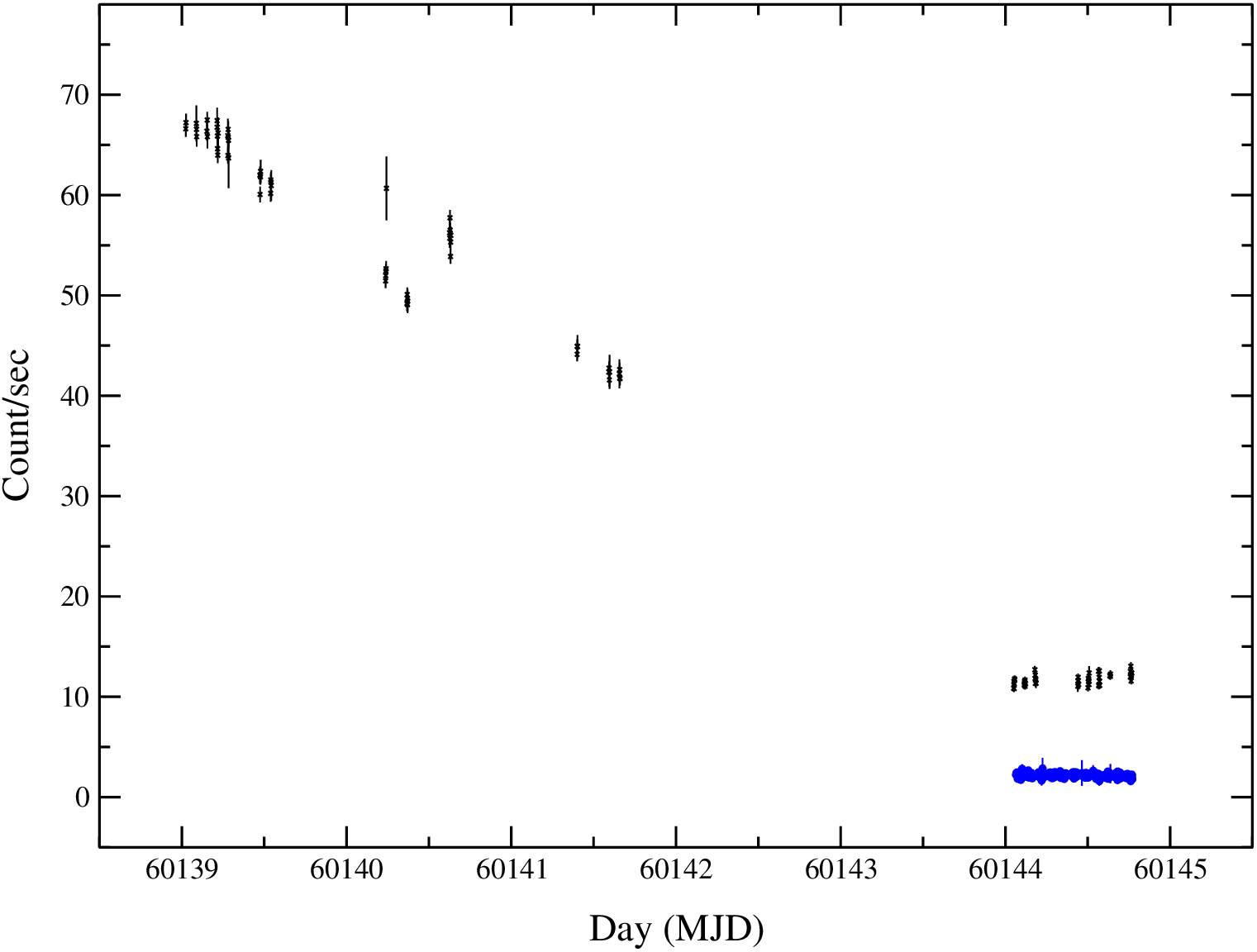}
    \caption{The 100~sec time binned light curves of {\it NICER} and {\it NuSTAR} data of the 2023 outburst. The black crosses represent the light curves ($0.5-10$~keV) from four {\it NICER} observations. The blue dots represent the $3-78$~keV {\it NuSTAR} light curve. The background-subtracted {\it NuSTAR} light curves from the two modules (FPMA and FPMB) were combined with {\tt lcmath} task.}
    \label{fig:nicer_nustar_lc}
\end{figure}

\subsection{Timing analysis}
During our studied period (from July 14 $-$ July 19) of the outburst in 2023, we noticed that the source count rate decreased consistently. In Fig.\ref{fig:nicer_nustar_lc}, we show the $0.5-10$~keV light curve during the observation period. We also show the $3-78$~keV light curve of {\it NuSTAR} (online blue curve) on July 19, 2023. 
In the case of {\it NuSTAR} light curves, the soft energy band count rate is approximately twice the hard-band count. No variation can be seen in any of the soft or hard band light curves.

\begin{figure*}
	\includegraphics[width=8.5cm]{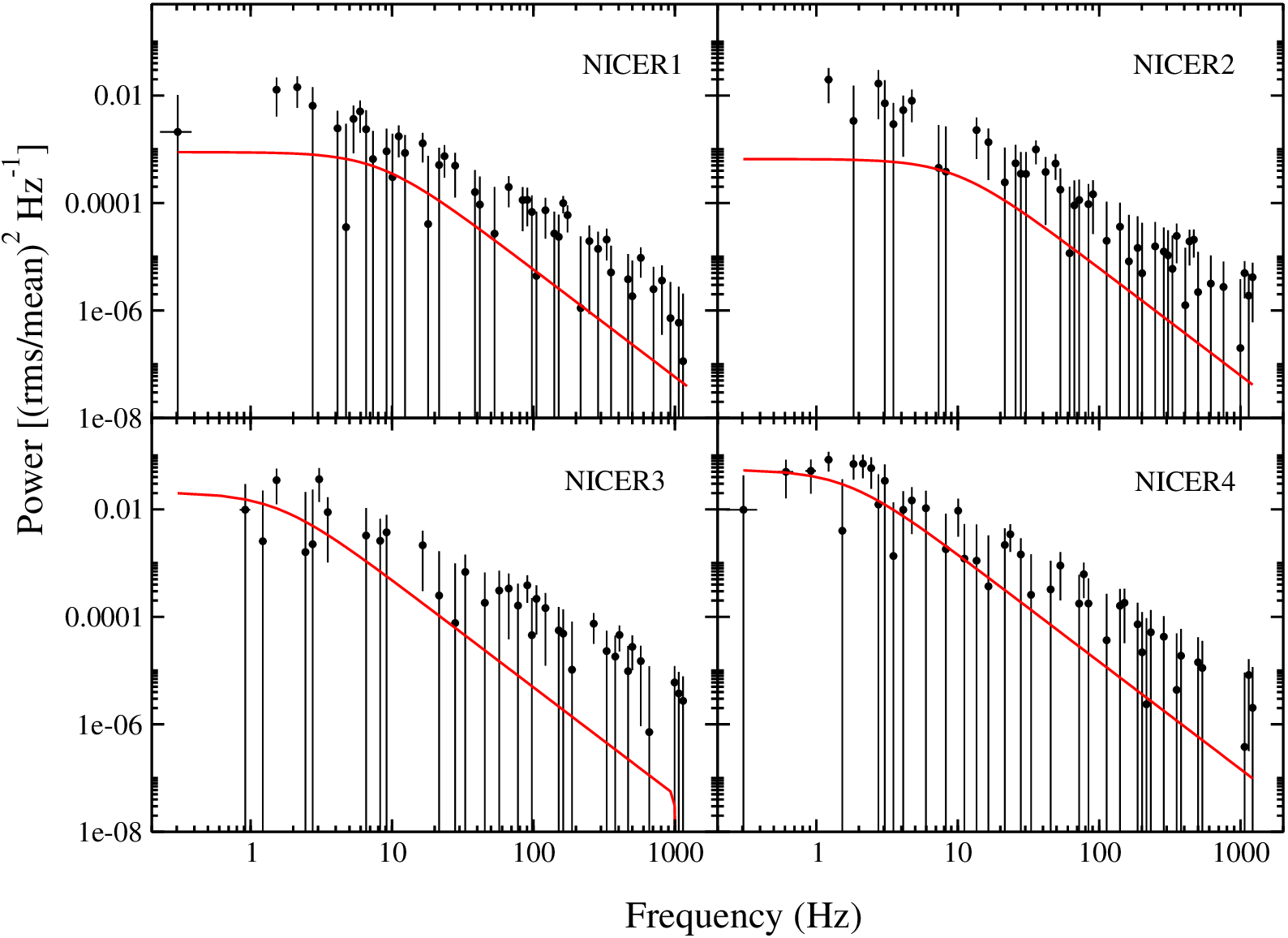}
    \includegraphics[width=8.2cm]{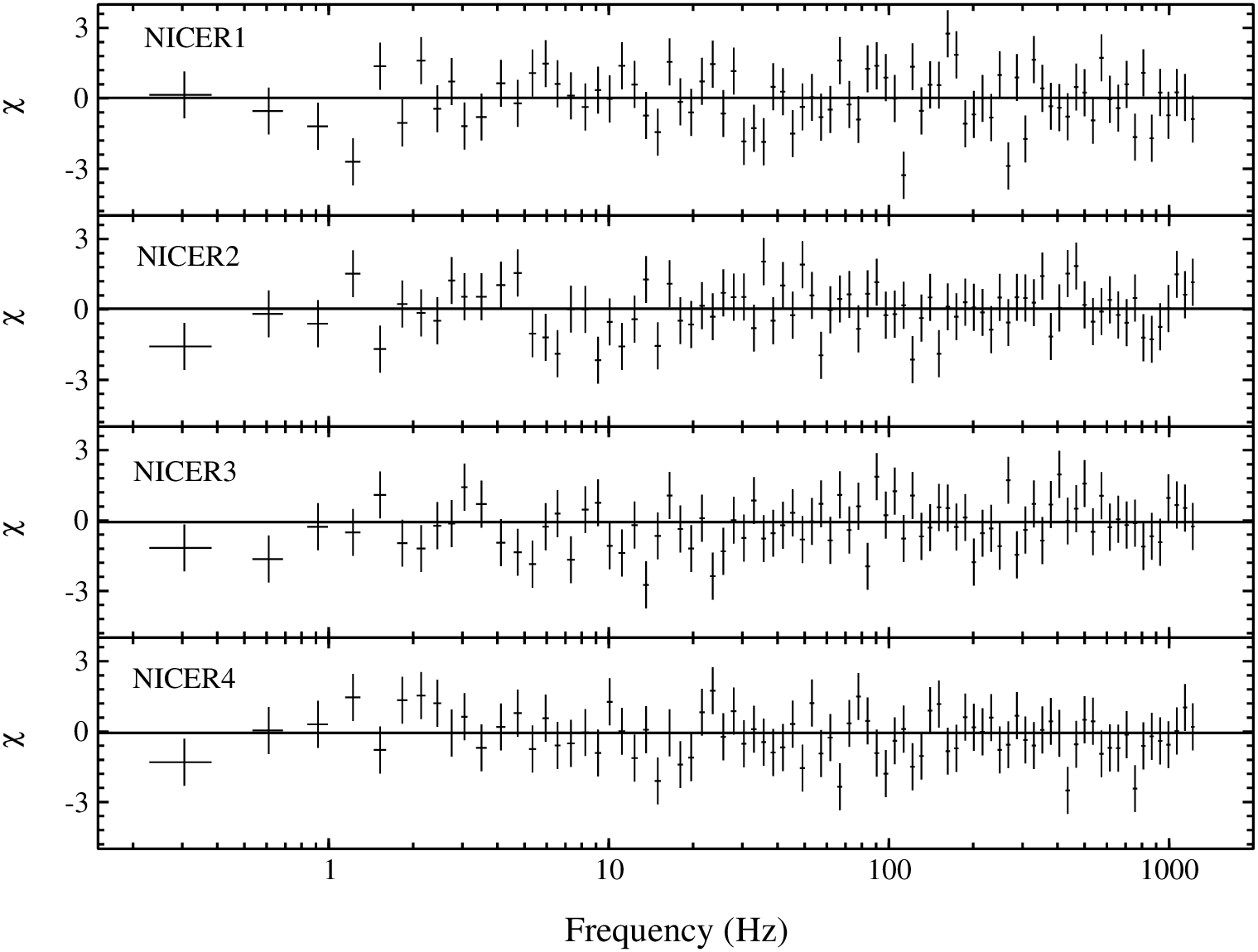}
    \caption{Left: Power density spectra (PDSs) of four {\it NICER} observations. Black dots and solid red lines represent the data and the Lorentzian model. Right: Residuals of the corresponding PDSs. The PDSs are created from $400~\mu s$ binned $0.5-10$~keV light curves with geometrical rebinning of 0.07 in {\tt xronos}.}
    \label{fig:nicer_pds}
\end{figure*}

\begin{figure}
	\includegraphics[width=8.5cm]{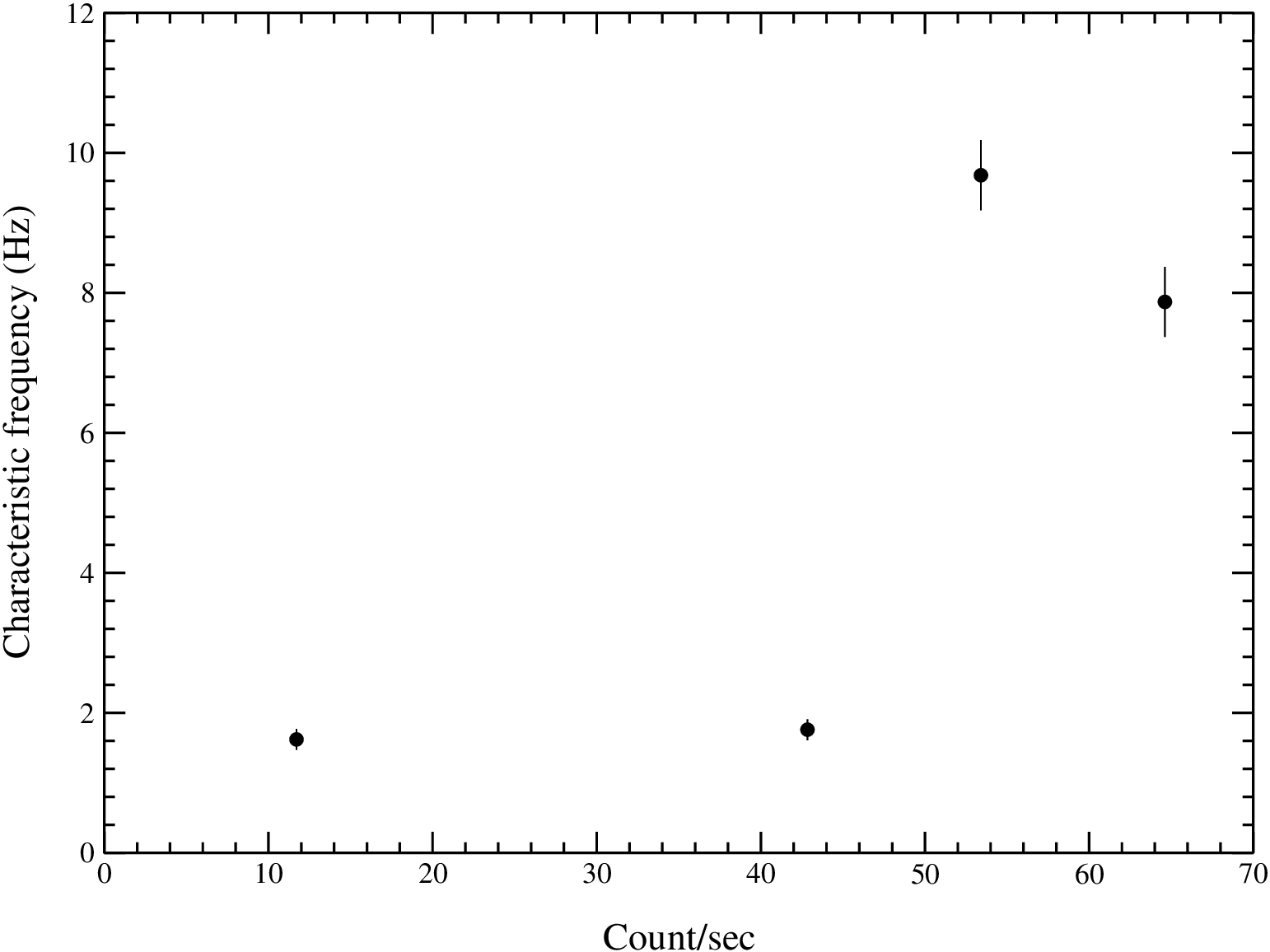}
    \caption{Variation of Characteristic frequency ($\nu_c$) with {\it NICER} count rate (Count/sec) for four observations.}
    \label{fig:chrfrq_count}
\end{figure}

To study the variability, we study the PDSs of the four {\it NICER} light curves generated from 400~$\mu$s light curves. These high-resolution light curves allowed us to investigate the QPO feature in the PDSs up to 1250~Hz. The PDSs do not show any QPO features. We fit the PDSs with Lorentzian models. The four PDSs and their corresponding residuals are presented in Fig.~\ref{fig:nicer_pds}. The zero-centered Lorentzian model is shown with the red curves in Fig.~\ref{fig:nicer_pds}. We calculate the characteristic frequency ($\nu_c$) of the broad-band noise of those PDSs in the $0.1-1250$~Hz frequency range. The characteristic frequency ($\nu_c$) is the frequency where the component contributes the most of its variance per frequency. If the centroid frequency is $\nu_0$ and the full-width at half-maximum is $\Delta\nu$, then the characteristic frequency ($\nu_c$) is $\sqrt{\nu_0+(\Delta\nu/2)^2}$ \citep{Nowak2000,Belloni2002}. The variation of $\nu_c$ with {\it NICER} count rate is shown in Fig.~\ref{fig:chrfrq_count}.

\begin{figure*}
\centering
	\includegraphics[width=7.5cm]{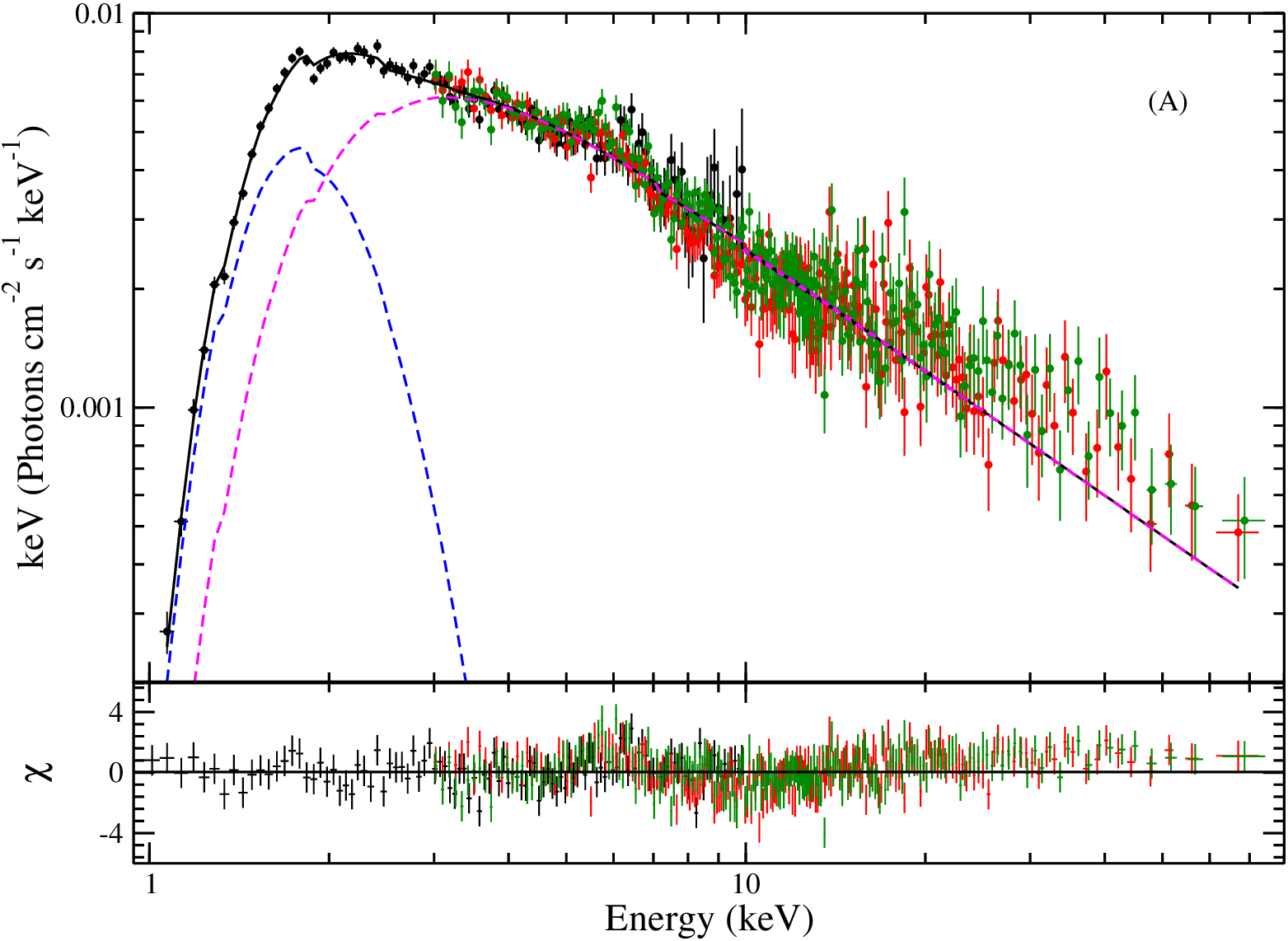}
 	\includegraphics[width=7.5cm]{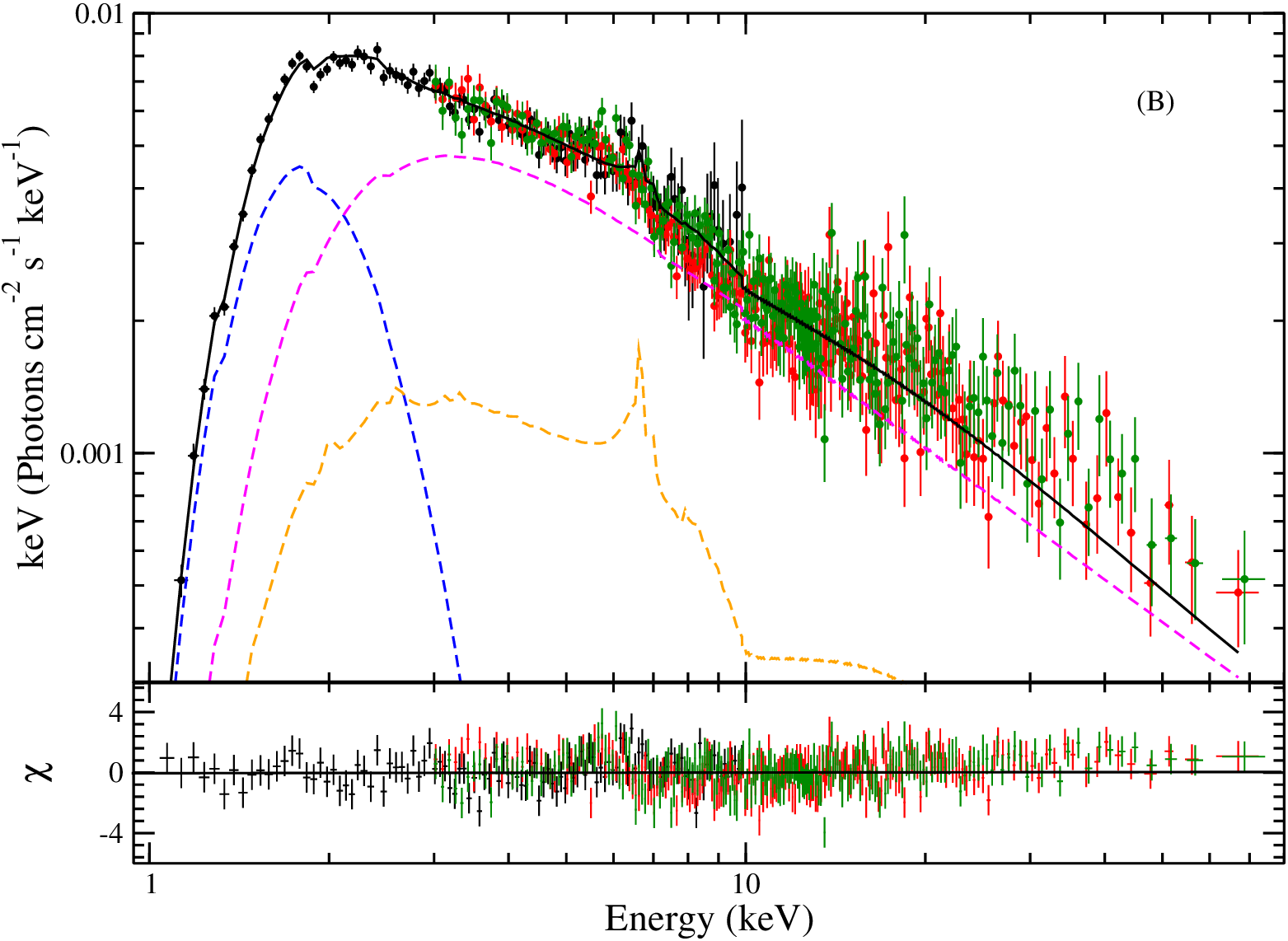}
    \includegraphics[width=7.5cm]{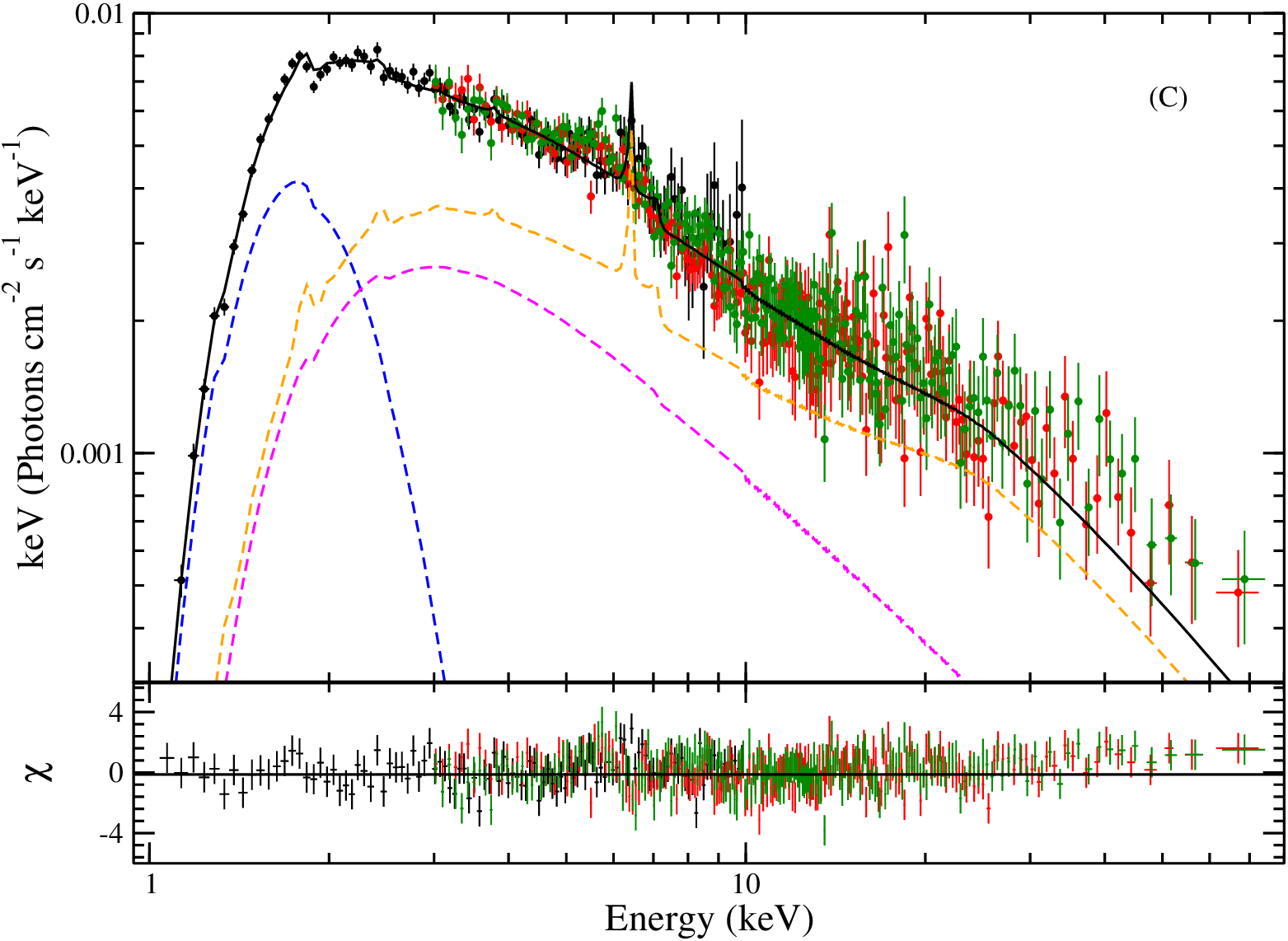}
    \includegraphics[width=7.5cm]{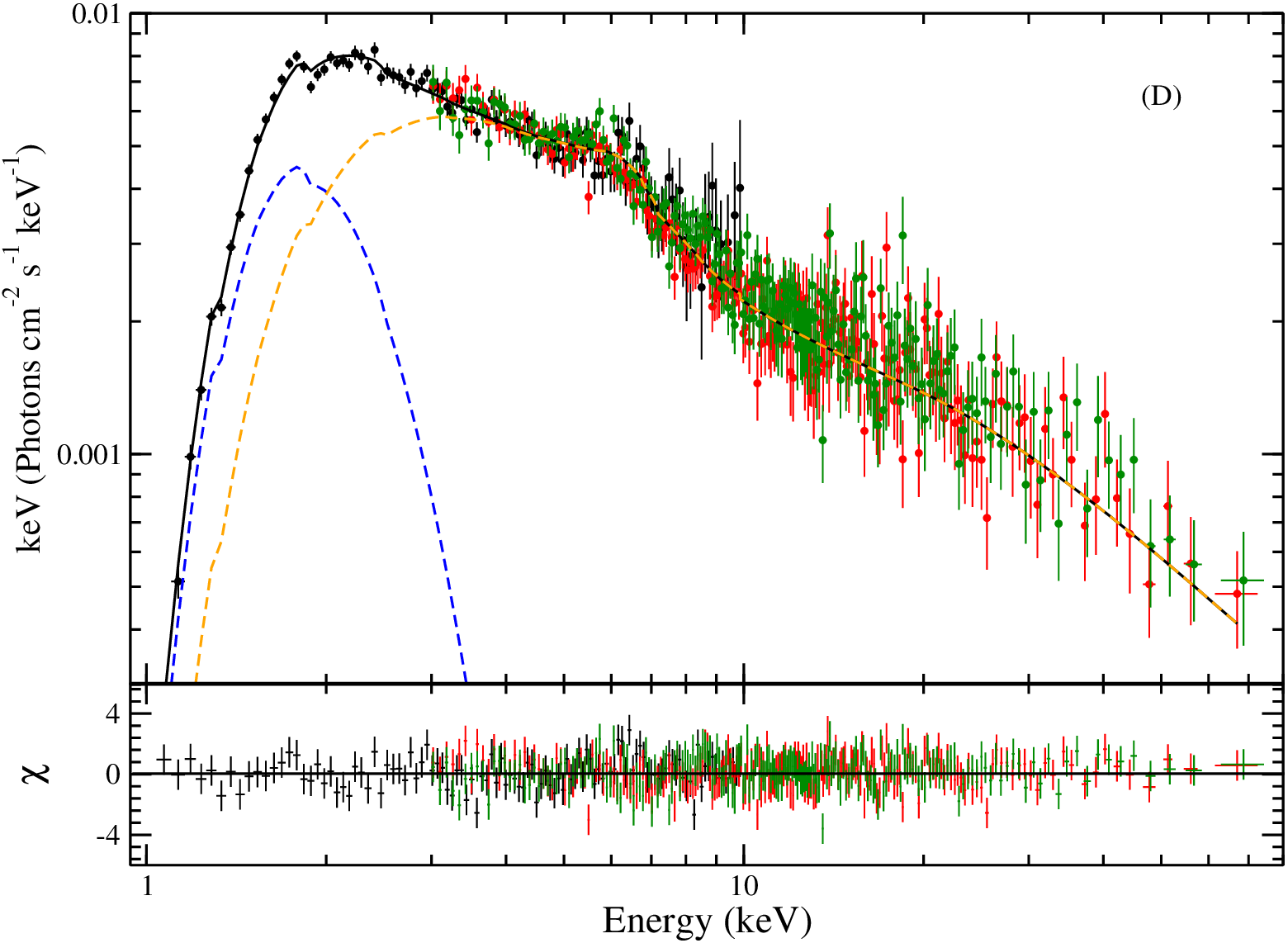}
    \caption{Model fitted combined NICER4 (1-10 keV) and NuSTAR (3-78 keV) spectrum with their corresponding residuals. Black, red and green represent the data of {\it NICER}, {\it NuSTAR} FPMA and FPMB, respectively. The lower panels represent the $\chi^2$ variation for fitted models. A, B, C and D corresponds to \textsc{tbabs*(DiskBB+PL)}, \textsc{Tbabs(DiskBB+PL+xillver)}, \textsc{Tbabs(DiskBB+ThComp~$\otimes$~DiskBB+xillver)}, and \textsc{tbabs*(DiskBB+relxill)} spectra respectively. The total model spectra are solid black lines in all panels (A, B, C, D). Dashed blue lines are the \textsc{DiskBB} components. \textsc{xillver} component is shown by orange dotted lines in panels B and C. The magenta dashed lines represent the \textsc{PL} components in panels A and B. In panel C, magenta \textsc{ThComp$\otimes$DiskBB} component. In panel D, the orange dashed line represents the \textsc{relxill} component.}
    \label{fig:comb_spectra}
\end{figure*}

\subsection{Spectral analysis}
We study the 2013 outburst of the source using combined NICER4 and NuSTAR on July 19, 2023. We also fit the four {\it NICER} data on July 14, July 15, and July 16 observations. The $1-10$~keV energy range of NICER4 and $3-78$~keV energy range of NuSTAR are selected. We use \textsc{xspec} \citep{Arnaud1996} for the spectral fit. 

We use a combination of phenomenological and physical models to fit the spectra. We use disk blackbody (\textsc{DiskBB}) and power-law (\textsc{PL}) models to obtain an overview of the accretion properties. The \textsc{DiskBB} model is used to describe the thermal emission from the accretion disk around a BH or NS. It is based on the multi-color disk blackbody model, which assumes the disk consists of multiple concentric annuli, each emitting as a blackbody with a different temperature \citep{Mitsuda1984}. The input parameters of this model include the inner disk temperature ($T_{\rm in}$), which represents the temperature at the inner disk radius, and the normalization ($Norm_{\rm DiskBB}$), which is proportional to the square of the inner disk radius ($R_{\rm in}$) and inversely proportional to the distance to the source (in 10~kpc unit). The normalization also depends on the inclination angle of the disk ($\cos\theta$).

Later, the spectra are fitted with more physical models (e.g., \textsc{ThComp} and \textsc{relxill}). \textsc{ThComp} \citep{Zdziarski2020} is an improved version of \textsc{nthComp} model \citep{Zdziarski1996} with actual Monte Carlo spectra from Comptonization \citep{Niedzwiecki2019}. The \textsc{ThComp} model characterizes spectra resulting from Comptonization by thermal electrons emanating from a spherical source. These electrons interact with seed photons distributed sinusoidally, similar to the \textsc{CompST} model \citep{Sunyaev&Titarchuk1980}. \textsc{ThComp} operates as a convolution model, enabling the Comptonization of various seed photon distributions, whether they are hard or soft. It adeptly accounts for both the up-scattering and down-scattering processes. In scenarios where certain seed photons, such as those from a blackbody or a disk blackbody, are up-scattered, \textsc{ThComp} offers a superior representation of the continuum shape resulting from thermal Comptonization compared to an exponentially cutoff power law. \textsc{ThComp} exhibits a considerably sharper cutoff compared to the exponential model. Additionally, the model furnishes an accurate portrayal of Comptonized spectra at energies that are in the same range as those of the seed photons. This model does not have a normalization parameter since its normalization follows from that of the seed photons.

\textsc{relxill} model combines the \textsc{relconv} relativistic convolution model with the \textsc{xillver} reflection model, effectively accounting for the relativistic effects and the ionization state of the accretion disk material \citep{Garcia2014, Dauser2014}. Key parameters of the \textsc{relxill} model include the emissivity indices (Index1, Index2), which describe how the intensity of the reflected emission falls off with distance from the BH. The spin parameter ($a$) quantifies the dimensionless spin of the BH, ranging from `0' (non-rotating) to `1' (maximally rotating). The inclination angle ($\theta$) specifies the angle of the accretion disk relative to the observer’s line of sight. The inner ($R_{\rm in}$) and outer ($R_{\rm out}$) disk radii define the spatial extent of the disk contributing to the reflection spectrum. The break radius ($R_{\rm br}$) is the boundary between $R_{\rm in}$ and $R_{\rm out}$ defined by the two emissivity indices. The photon index ($\Gamma$) characterizes the power-law shape of the primary X-ray source's spectrum, while the high-energy cutoff ($E_{\rm cut}$) indicates the energy at which the spectrum steepens, often related to the coronal temperature. The iron abundance ($A_{\rm Fe}$) and ionization parameter ($\xi$) influence the strength and shape of the reflection features, particularly the Fe~K$\alpha$ line. The reflection fraction ($f_{\rm refl}$) specifies the proportion of the reflection component relative to the direct power-law emission. The model's normalization ($Norm_{\rm relxill}$) scales the overall flux of the reflected component \citep{Dauser2016}.

We use \textsc{vern} cross-section \citep{Verner1996} and \textsc{wilm}'s abundance \citep{Wilms2000} for the spectral fit. The \textsc{constant} parameter is used for combining the three spectra of {\it NICER}, {\it NuSTAR}/FPMA and {\it NuSTAR}/FPMB on July 19 observations. We fixed the value of the \textsc{constant} to 1 for {\it NICER} spectra and kept it free for FPMA and FPMB spectra. In this way, we get the same normalization value of the model parameters. The parameters for the different combinations of models are given in Table~\ref{tab:table2} and Table~\ref{tab:table3}. The errors are calculated using \textsc{xspec}'s {\tt fit err} command with 90\% confidence.

\subsubsection{Model 1}
First, we fit the combined spectra with the absorbed multi-color \textsc{DiskBB}\citep{Mitsuda1984,Makishima1986} and \textsc{PL} model. We use \textsc{Tbabs} for the absorption component to find the neutral hydrogen column density along the line of sight. This parameter is kept free during the fit. The total model can be represented as: \textsc{Tbabs*(DiskBB+PL)}. A \textsc{gabs} model is used to incorporate the prominent Si edge feature around 1.8 keV. 
The obtained parameter values are line energy $\sim$ 1.90$\pm$0.02 keV, width $\sim0.03\pm0.01$ keV, and strength $\sim0.03\pm0.01$. The parameters of \textsc{gabs} are then fixed during the analysis with other combinations of models. We obtain the hydrogen column density ($N_{\rm H}$) for the best-fit spectra for the four epochs as $4.3\pm0.1~\times10^{22}$ ${\rm cm}^{-2}$. The inner disk temperature ($T_{\rm in}$) are $0.57\pm0.01$, $0.51\pm0.01$, $0.48\pm0.01$, and $0.28\pm0.01$~keV. The normalization of \textsc{DiskBB} model ($Norm_{\rm DiskBB}$) of these four epochs are $820\pm40$, $980\pm100$, $1110\pm80$, and $3900\pm900$ respectively. For the NICER1 observation, we obtained $\Gamma$=0.2, which is unphysical. Hence, we fixed it at 2.5 following the fit from the next observation on July 15. The NICER2, NICER3, and NICER4+NuSTAR give $\Gamma$ values of $2.50\pm1.00$, $2.30\pm0.40$ and $2.05\pm0.01$, respectively. The best-fit spectrum (with $\chi^2_{\rm red}\sim1.35$) of July 19 is given in Fig.~\ref{fig:comb_spectra}A.

\subsubsection{Model 2}
In the case of the first three {\it NICER} observations, we do not see any significant contribution of Fe~K$\alpha$ line around 6~keV. Although, for combined spectra on July 19, we see a hump around 6 keV (see Fig.~\ref{fig:comb_spectra}A). To consider the iron emission line contribution around 6 keV we add \textsc{Gaussian} component. We do not add the \textsc{Gaussian} to the best-fit {\it NICER} observations because the contribution is insignificant. We obtain a very broad line width ($\sigma\sim$1.51 keV) for the \textsc{Gaussian} with a line energy of $<5$~keV. Hence, we replace the \textsc{Gaussian} with \textsc{xillver} model \citep{Garcia2013,Dauser2013}. The \textsc{Tbabs(DiskBB+PL+xillver)} best-fit spectrum is shown in the Fig.~\ref{fig:comb_spectra}B. The fitted model parameters are given in Table~\ref{tab:table3}. We obtain an iron abundance of 1.0$^{+0.5}_{-0.1}$ solar value. The ionization parameter retains a value of 3.3$\pm$0.1. The inclination angle ($\theta$) obtained is 20$^\circ$$^{+16}_{-12}$. With a free high-energy cut parameter, the best fit is obtained at 1000 keV, which is well outside the NuSTAR energy range. So, we fixed the high energy cut to 1000 keV in the next study. The reflection fraction is fixed at -1, to consider only the contribution from the line emission. This fit gives a $\chi^2_{\rm red}$ of $\sim$ 1.07.

\subsubsection{Model 3}
Next, we replace the phenomenological \textsc{PL} model (see, Model 1) with a more physical \textsc{ThComp} model \citep{Zdziarski2020}. We fit the spectra with \textsc{Tbabs(DiskBB+ThComp~$\otimes$~DiskBB)}. We set the energy range from 0.01 to 1000.0, as the model requires, to give the correct result. The convolution model's $T_{\rm in}$ is set to the value of the \textsc{DiskBB} model temperature. \textsc{xillver} model is also included for the NICER4+NuSTAR spectrum fit. 
Since, for the case of the first three {\it NICER} observations, the energy range considered is $1-10$~keV, we fixed the $k{\rm T_e}$ to 50~keV. We noticed that the $\Gamma$ and $T_{\rm in}$ values retain similar results as Model 1 for these spectra. The $\Gamma$ for the combined spectra is $2.07\pm0.01$. The \textsc{xillver} model parameters for the combined spectra are $A_{\rm Fe}\sim4.0^{+0.6}_{-0.4}$, log~$\xi\sim1.7\pm0.1$, $\theta\sim27^\circ$$^{+14}_{-18}$. The best-fit spectrum of July 19 is given in Fig.~\ref{fig:comb_spectra}C. We present the combined spectral fitted results of the NICER4+NuSTAR observation in Table~\ref{tab:table3}.

\subsubsection{Model 4}
We also fit the July 19's combined spectra with \textsc{tbabs(Diskbb+relxill)} model. Since the first three spectra do not show any significant iron emission line features and no reflection features can be seen, we have not fit these three {\it NICER} observations with this model combinations. The \textsc{DiskBB} model gives similar results to the previous combinations with $T_{\rm in}\sim0.32$ keV. We fixed the emissivity indices (Index1 and Index2) to 3, to Newtonian emissivity. The outer radius ($R_{\rm out}$) is fixed at 400~$r_g$. Increasing the $R_{\rm out}$ does not change the results. Fig.~\ref{fig:comb_spectra}D represents the best-fit spectrum. We obtain the best-fit model parameters to be spin (a) $\sim0.98\pm0.01$, $\theta$ of the inner accretion disk $\sim18^{+5}_{-7}$, inner accretion disk radius (R$_{\rm in}$) $\sim4.5^{+1.0}_{-2.0}$ $r_{\rm ISCO}$ ($r_{\rm ISCO}$ is the inner stable circular orbit's radius), $\Gamma\sim1.94\pm0.02$. The $A_{\rm Fe}$ and log$\xi$ for the best fit are $\sim5.5\pm0.4$ and $\sim3.4^{+0.7}_{-0.1}$ respectively. The best fit gives a $\chi^2_{\rm red}$ value of $\sim$0.98. The best-fit parameters are given in Table~\ref{tab:table3}.

From the $\chi^2/dof$ values of the different combinations of models, we conclude that for the NICER1-3 observations Model 1 gives the best-fit. For combined NICER4 and NuSTAR observation, Model 4 is the better model for explaining the spectra with fewer (physical) parameters and satisfactory statistics.

\begin{table*}
    \centering
    \caption{Spectral Results for three NICER observations.}
    \label{tab:table2}
    \begin{tabular}{lccr} 
        \hline
        DiskBB+PL (Model 1) & NICER1 & NICER2 & NICER3 \\
        Parameters & 14/07/2023 & 15/07/2023 & 16/07/2023 \\
        \hline
        N$_{\rm H}$ ($\times10^{22}~{\rm cm}^{-2}$) & 4.3$\pm$0.1 & 4.3$\pm$0.1 & 4.3$\pm$0.1 \\
        T$_{\rm in}$ (keV) & 0.57$\pm$0.01 & 0.52$\pm$0.01 & 0.48$\pm$0.01 \\
        Norm$_{\rm DiskBB}$ & 820$\pm$40 & 980$\pm$100 & 1110$\pm$80 \\
        $\Gamma$ & $2.5^*$ & 2.50$\pm$1.00 & 2.30$\pm$0.40 \\
        Norm$_{\rm PL}$ ($\times10^{-2}$) & 4.0$\pm$0.3 & 3.0$\pm$0.3 & 2.0$\pm$1.0 \\
        $\chi^2/dof$ & $\sim$1.68 & $\sim$1.03 & $\sim$1.00 \\
        \hline
    \end{tabular}
    
    \noindent{The best-fit parameters of the three {\it NICER} observations using \textsc{DiskBB+PL} model. The errors are calculated using {\tt fit err} command and represent 90\% confidence level. The `$^*$' denotes the frozen/fixed parameters. The first column represents the parameters of the respective models. Absorption models \textsc{Tbabs} is used for all model combinations as a multiplicative model.}
\end{table*}

\begin{table*}
	\small
 \centering
	\caption{Spectral Results for the combined NICER4+NuSTAR spectrum on 19/07/2023.}
	\label{tab:table3}
	\begin{tabular}{lcccr} 
 
		\hline
  Model/& Model 1 & Model 2 & Model 3 & Model 4 \\
  Parameters & (\textsc{DiskBB+PL}) & (\textsc{DiskBB+PL} & (\textsc{DiskBB+xillver} & (\textsc{DiskBB}\\
  & & \textsc{+xillver})& \textsc{+ThComp$\otimes$DiskBB}) & \textsc{+relxill})\\
  \hline
\multicolumn{3}{l}{\textsc{TBabs}}\\
\cmidrule{1-1}
  $N_{\rm H}$ ($\times10^{22}~{\rm cm}^{-2}$) & 4.5$\pm$0.2 & 4.3$\pm$0.2 & 4.0$\pm$0.1 & 4.3$\pm$0.1 \\
  &&&&\\
  \multicolumn{3}{l}{\textsc{DiskBB}}\\
  \cmidrule{1-1}
  $T_{\rm in}$ (keV) & 0.28$\pm$0.01 & 0.30$\pm$0.01 & 0.33$\pm$0.01 & 0.32$\pm$0.01 \\
  $Norm_{\rm DiskBB}$  &  3900$\pm$900 & 2500$^{+1200}_{-800}$ & 7700$^{+3700}_{-3400}$ & 1500$^{+400}_{-300}$ \\
  &&&&\\
  \multicolumn{3}{l}{\textsc{PL}}\\
  \cmidrule{1-1}
  $\Gamma$      & 2.05$\pm$0.01 & 2.00$\pm$0.02 & - & - \\
  $Norm_{\rm PL}$ ($\times10^{-3}$) & 30$\pm$1 & 21$\pm$1 & - & - \\
  &&&&\\
  \multicolumn{3}{l}{\textsc{ThComp}}\\
  \cmidrule{1-1}
 $kT_{\rm e}$ (keV)  & - & - & 149$\pm$0.4 & - \\
 $\Gamma$      & - & - & 2.07$\pm$0.01 & - \\
 $f_{\rm cov}$ & - & - & 1.00$^{+0.05}_{-0.01}$ & - \\
 $Norm_{\rm ThComp}$ & - & - & 2350$^{+360}_{-370}$ & - \\
 &&&&\\
 \multicolumn{3}{l}{\textsc{xillver}}\\
 \cmidrule{1-1}
 $A_{\rm Fe}$ (in $Solar$) & - & 1.0$^{+0.5}_{-0.1}$ & 4.0$^{+0.6}_{-0.4}$ & -\\
 $E_{\rm cut}$ (keV) & - & 1000$^*$ & 1000$^*$ & - \\
 log~$\xi$ (erg~cm~s$^{-1}$) & - & 3.3$^{+0.1}_{-0.1}$ & 1.7$^{+0.1}_{-0.1}$ & - \\
 Inclination ($\theta^\circ$) & - & 20$^{+16}_{-12}$ & 27$^{+14}_{-18}$ & - \\
 $Norm_{\rm xillver}$ ($\times10^{-4}$) & - & 2.0$\pm$0.2 & 73$\pm$5 & - \\
 &&&&\\
 \multicolumn{3}{l}{\textsc{relxill}}\\
 \cmidrule{1-1}
 Index1 & - & - & - & 3$^*$ \\
 Index2 & - & - & - & 3$^*$ \\
 $a$ & - & - & - & 0.98$\pm$0.01 \\
 Inclination ($\theta^\circ$) & - & - & - & 18$^{+5}_{-7}$ \\
 $R_{\rm in}$ ($r_{\rm ISCO}$) & - & - & - & 4.5$^{+1.0}_{-2.0}$ \\
 $R_{\rm out}$ (GM/c$^2$) & - & - & - & 400$^*$ \\
 $\Gamma$ & - & - & - & 1.94$\pm$0.02 \\
 $A_{\rm Fe}$ (in $Solar$) & - & - & - & 5.5$\pm$0.4 \\
 $E_{\rm cut}$ (keV) & - & - & - & 1000$^*$ \\
 log$\xi$ (erg~cm~s$^{-1}$) & - & - & - & 3.4$^{+0.7}_{-0.1}$ \\
 $f_{\rm refl}$ & - & - & - & 0.54$\pm$0.01 \\
 $Norm_{\rm relxill}$ ($\times10^{-6}$) & - & - & - & 332$\pm$1 \\
 $\chi^2/dof$ & 1304.42/964 & 1035.14/959 & 1178.16/955 & 942.37/957 \\
 & $\sim$1.35 & $\sim$1.07 & $\sim$1.23 & $\sim$0.98\\

  \hline
  
   \hline
      \end{tabular}
      
\noindent{The superscripts and subscripts represent the positive and negative errors, respectively, and are calculated using {\tt fit err} command. The errors are estimated at 90\% confidence level. The `$^*$' denotes the frozen/fixed parameters. The first column represents the parameters of the respective models. The combinations of models are given in the first row. The individual models are also given in the first row before the parameters of the model are given. Absorption models \textsc{Tbabs} is used for all model combinations as a multiplicative model.}
\end{table*}

\section{Discussions}
We study the accretion properties of XTE~J1856+053 during its 2023 outburst using four {\it NICER} and one {\it NuSTAR} data from July 14 to July 19, 2023. The timing properties are investigated with high resolution {\it NICER} data to search for the QPOs during the outburst period. For the spectral study with different models, the $1-10$~keV NICER4 spectrum is combined with the $3-78$~keV NuSTAR spectrum. We also use three other {\it NICER} data for spectral study.

\subsection{PDS with no QPO features}
We noticed that the source count rate decreased consistently from the start of our observation period. The nature of the compact X-ray object is not confirmed via any dynamical mass estimation. The spectral study of the 2007 outburst using {\it XMM-Newton} observation suggested it to be a BH of mass $1.3-4.2~M_\odot$ \citep{Sala2008}. No pulsations or thermonuclear bursts have been detected in any of the previous outbursts or in it's recent outburst, which is a clear signature of the NS binary. The kHz oscillation is a common feature for NS binaries \citep{disalvo2001,Bult2018,Mendez2021}, which is also absent in the PDSs of our studied observations. However, the absence of kHz oscillations does not totally discard the system to host an NS. We searched for low-frequency and high-frequency QPOs using 400~${\rm \mu s}$ light curves of {\it NICER}, but found none. The power of the PDS is decreased after $\sim$10~Hz. In general, the power of the PDSs of a stellar-mass BH decreases after 10-50 Hz \citep{Sunyaev2000}, showing a steep decreasing trend compared to NS binaries. The absence of the kHz oscillations and the decrease of the power after 10~Hz support it to be a potential BH candidate from the timing analysis.
We calculate the characteristic frequencies ($\nu_c$) for the four {\it NICER} observations from $0.1-1250$~Hz. The $\nu_c$ shows correlations with the count rate. Since, in the SS, the disk would be much closer to the BH, it shows more variability than in the HS. A positive correlation of the $\nu_c$ with the count rate thus indicates the evolution of the source from SS to HS. A similar correlation has been noticed in MAXI J0637-430 during the 2019 outburst along different spectral states \citep{Jana2021}.

\subsection{Spectral state transition}
We fit the spectra with different combinations of phenomenological and physical models to understand the accretion properties during the outburst. We use a total of four {\it NICER} spectra and one {\it NuSTAR} spectra for the spectral analysis. One of the {\it NICER} observation times overlapped with the {\it NuSTAR} observation, so we simultaneously analyzed the combined {\it NICER} and {\it NuSTAR} spectra. From the overall analysis with \textsc{DiskBB} and \textsc{PL} model of the four observations, we noticed that the $T_{\rm in}$ decreases consistently from 0.57 to 0.28 keV. Also, $\Gamma$ decreases during this period. A similar trend was observed during the transition from SS towards HS for other BHCs (see, \citealt{Remillard&McClintock2006,CadolleBel2004,Wang2018} and reference therein). The high values of $\Gamma$ ($\sim$2-2.5) indicate the source is in the SS during the first three observations. Also, the decreasing count rate (see Fig~\ref{fig:nicer_nustar_lc}) supports that the source is in the declining phase. The rising phase of the outburst was not detected, which is also observed in some outbursts of other BHs showing a fast-rising (missed) state (e.g., XTE J1755–324 in 1997; \citealt{Revnivtsev1998}, XTE~J1720-318 in 2003; \citealt{CadolleBel2004}, XTE J1726-476 in 2005; \citealt{Levine2005}). The $\Gamma$ value even decreases when we fit the last observation with the \textsc{relxill} model. The decreasing nature of the $\Gamma$ and $T_{\rm in}$ implies the source is transitioning from SS to IMS. Since we have only four observations (from July 14 to July 19), we could not confirm the exact state transition (whether in HIMS or SIMS on the last observation day). 

\subsection{Reflection and iron emission line}
We notice a prominent iron emission line feature in the {\it NuSTAR} spectrum. The emission line of iron is a result of the irradiation of the inner accretion disk by a hot X-ray source. The shape of the line is determined by the gravitational redshifts, light bending, and Doppler effects. Since the iron emission line originates from the innermost region of the accretion disk, it serves to probe the strong gravitational environment in the vicinity of the compact object. A Compton hump around 20-30 keV is also a reflection feature that is commonly observed in BH binary spectra. For our studied observations, the Compton hump is not significant in the spectra. The iron line profile is also not significant in the first three {\it NICER} observations, possibly because these three observations are in SS. The weak power-law component in the SS results in a weak irradiation of the accretion disk and, thus, a weak or none reflection component. The iron line emission is broad in the last observation while fitting with a \textsc{Gaussian} component. Similar properties are observed in BHC XTE J1908+094 during 2002 outburst. The Fe~K$\alpha$ line disappears in the SS due to the weak PL component \citep{Gogus2004}. The best fit with \textsc{xillver} or \textsc{relxill} models give high ionization parameter values as well as a double-peaked profile of the emission line. This feature also suggests the source is in the IMS, and the inner accretion disk is very close to the BH. The high ionization parameters implied that the irradiation is high and the emission lines come from a larger optical depth (Thompson optical depth) and get thermally broadened \citep{Fabian&Ross2010}.

\subsection{Inner accretion properties and Spin}
From the spectral analysis, we notice a decreasing inner disk temperature, which means the inner disk radius increases during our studied observation period. We obtain the inner disk radius is $4.5^{+1.0}_{-2.0}$ $r_{\rm ISCO}$ on the last observation day from the \textsc{relxill} model. During the whole outburst, the spectra are dominated by the thermal component. The apparent inner accretion disk radius ($r_{\rm in}$) can be estimated from the DiskBB model normalization parameter ($Norm_{\rm DiskBB}$) as
\begin{equation}
    r_{\rm in} ({\rm km})=\sqrt{\frac{Norm_{\rm DiskBB}}{\cos\theta}}d_{10},
\end{equation}
where $\theta$ is the inclination of the accretion disk and $d_{10}$ is the distance in 10~kpc. To correct the spectral hardening \citep{Shimura-Takahara1995} and inner boundary correction \citep{Kubota1998}, we consider a constant spectral hardening factor $\kappa=1.7$ and inner boundary correction factor $\xi\sim0.41$. The corrected inner radius is, thus,
\begin{equation}
    R_{\rm in} ({\rm km}) = \kappa^2 \xi r_{\rm in} \simeq \frac{1.18\sqrt{Norm_{\rm DiskBB}}}{\sqrt{\cos\theta}}d_{10}.
\end{equation}
Assuming a disk inclination ($\theta$) of 18$^\circ$, and 10~kpc distance, we obtain $R_{\rm in}$ to vary from 35$\pm$1 km, 38$\pm$2 km, 40$\pm$1 km, and 47$\pm$5 km respectively during the four observations. 

One of the crucial parameters of a BH is spin, which can also be modeled by the iron line emission profile \citep{Fabian2012}. The broadening of the line profile gives the gravitational redshift resulting from the inner accretion disk. The spin can be estimated from the inner radius compared to the ISCO ($r_{\rm ISCO}$). The $r_{\rm ISCO}$ is a monotonic function of dimensionless spin parameters. This method of spin calculation is independent of the mass of the BH and distance to the system. We fit the spectra with angle-dependent model \textsc{relxill} to obtain the spin parameter. The obtained BH spin is $>0.9$, implying a maximally rotating Kerr BH. We also obtain the inclination angle of the inner accretion disk to be $\theta\sim18^\circ$$^{+5}_{-7}$.

We also obtain the inner accretion radius, $\sim4.5\pm1.5~r_{\rm ISCO}$ from Model 4. We estimate the $R_{\rm in}$ from $Norm_{\rm DiskBB}$ (See, Eq. 2). From the spectral fits of the two models, we can write,
\begin{equation}
    (4.5\pm1.5)~r_{\rm ISCO} \sim R_{\rm in},
\end{equation}
where, $r_{\rm ISCO}$ for a maximally rotating Kerr BH is considered to be between GM/c$^2$ and 9GM/c$^2$. For a Schwarzschild BH, $r_{\rm ISCO}$ is fixed at 6GM/c$^2$. We can estimate the mass of the BH ($M_{\rm BH}$) considering the BH to be a maximally rotating Kerr BH with prograde motion.
Considering 5~kpc, 10~kpc, and 15~kpc distances, we obtain the mass of the BH to be $4\pm1$ $M_\odot$, $7\pm2$ $M_\odot$, and $10\pm4$ $M_\odot$, respectively.

Although this result does not make any concrete conclusion regarding the $M_{\rm BH}$, it is evident that the mass value resides in a range of a stellar-mass BH candidate. Further successful optical observations in the quiescent state would enable the mass of the BH to be determined more accurately.

\section{Summary}
We investigated the accretion properties of the soft X-ray transient XTE J1856+053 during its 2023 outburst using {\it NICER} and {\it NuSTAR} data. Our analysis revealed several key findings:

\begin{itemize}
    \item We observed a consistent decrease in the source count rate throughout the observation period. We did not detect QPOs during our observation period.

    \item The spectral analysis indicated a transition from the Soft State (SS) to the Intermediate State (IMS) based on decreasing inner disk temperatures ($T_{\rm in}$) and power-law indices ($\Gamma$). However, due to limited observations, the exact state transition (HIMS or SIMS) couldn't be confirmed.

    \item We observed a prominent iron emission line in the {\it NuSTAR} spectrum, indicative of a strong gravitational environment near the compact object. The iron line profile suggested the source was in the IMS, with a close inner accretion disk.

    \item The decreasing inner disk temperature implied an increasing inner disk radius. The estimated BH spin parameter ($>0.9$) from the iron line profile suggested a maximally rotating Kerr BH. However, further observations are needed to determine the BH mass more precisely.
\end{itemize}

\begin{acknowledgments}
We would like to thank the anonymous referee for their constructive comments and suggestions, which have significantly improved the quality of this paper. D.C. and H.K.C acknowledge the grants NSPO-P-109221 of Taiwan Space Agency (TASA) and NSTC-112-2112-M-007-053 of National Science and
Technology Committee of Taiwan. 
AJ acknowledges support from Fondecyt postdoctoral fellowship (3230303). 
\end{acknowledgments}

\vspace{5mm}
\facilities{{\it NICER, NuSTAR}}


\software{HeaSoft \url{https://heasarc.gsfc.nasa.gov/docs/software/heasoft/},  }



\appendix


\begin{figure}[!ht]
\centering
	\includegraphics[width=7.5cm]{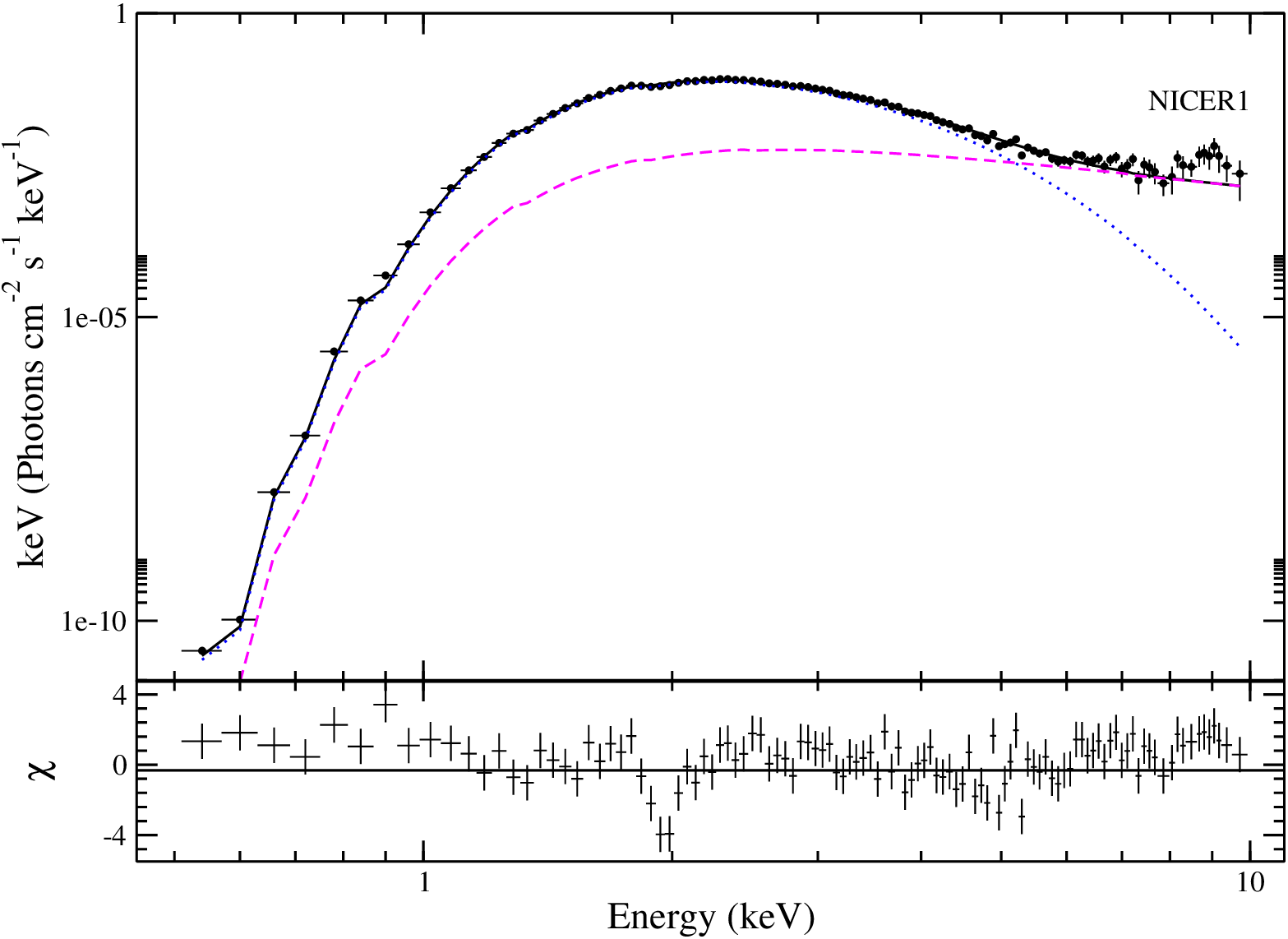}
 	\includegraphics[width=7.5cm]{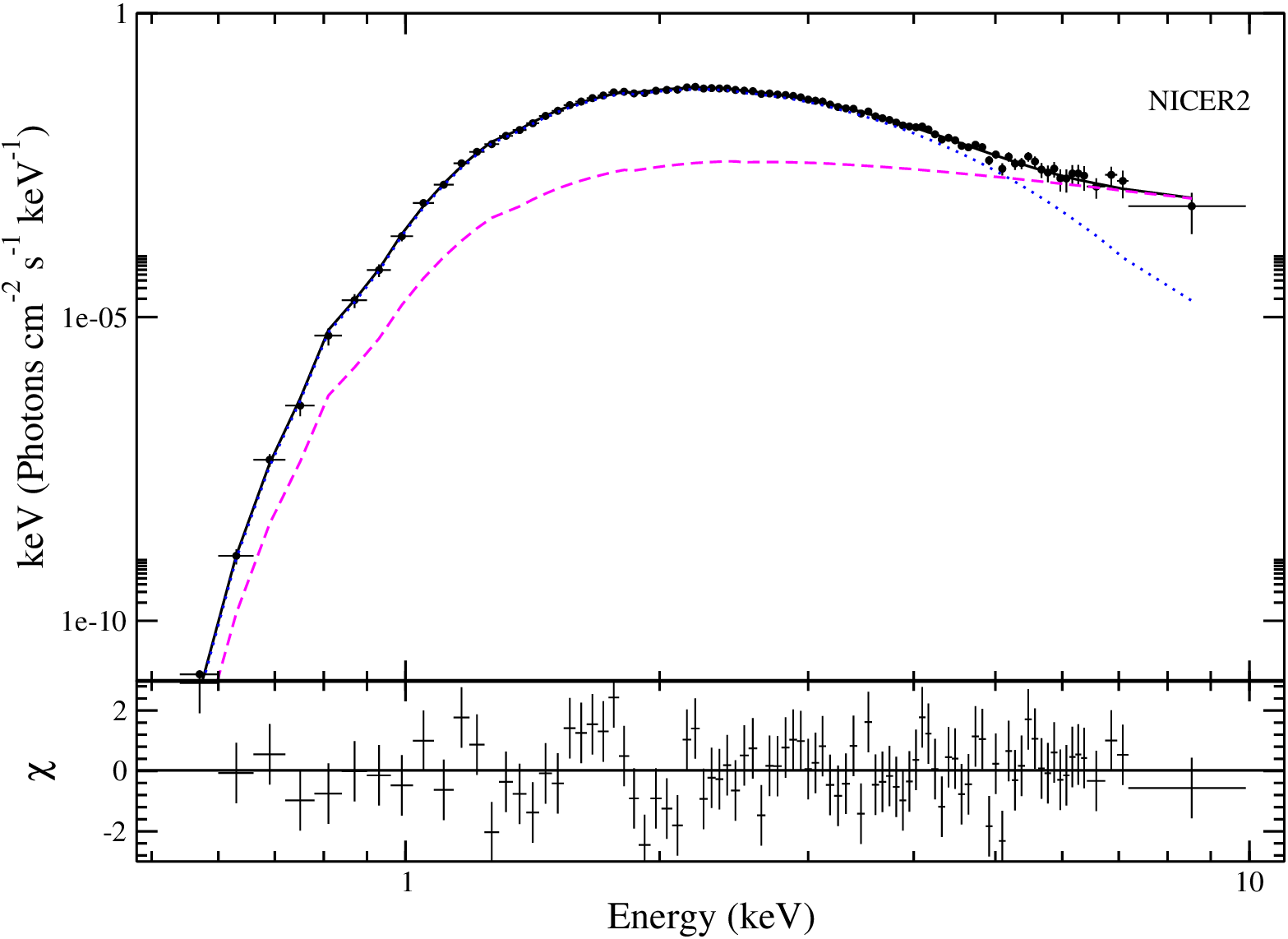}
    \includegraphics[width=7.5cm]{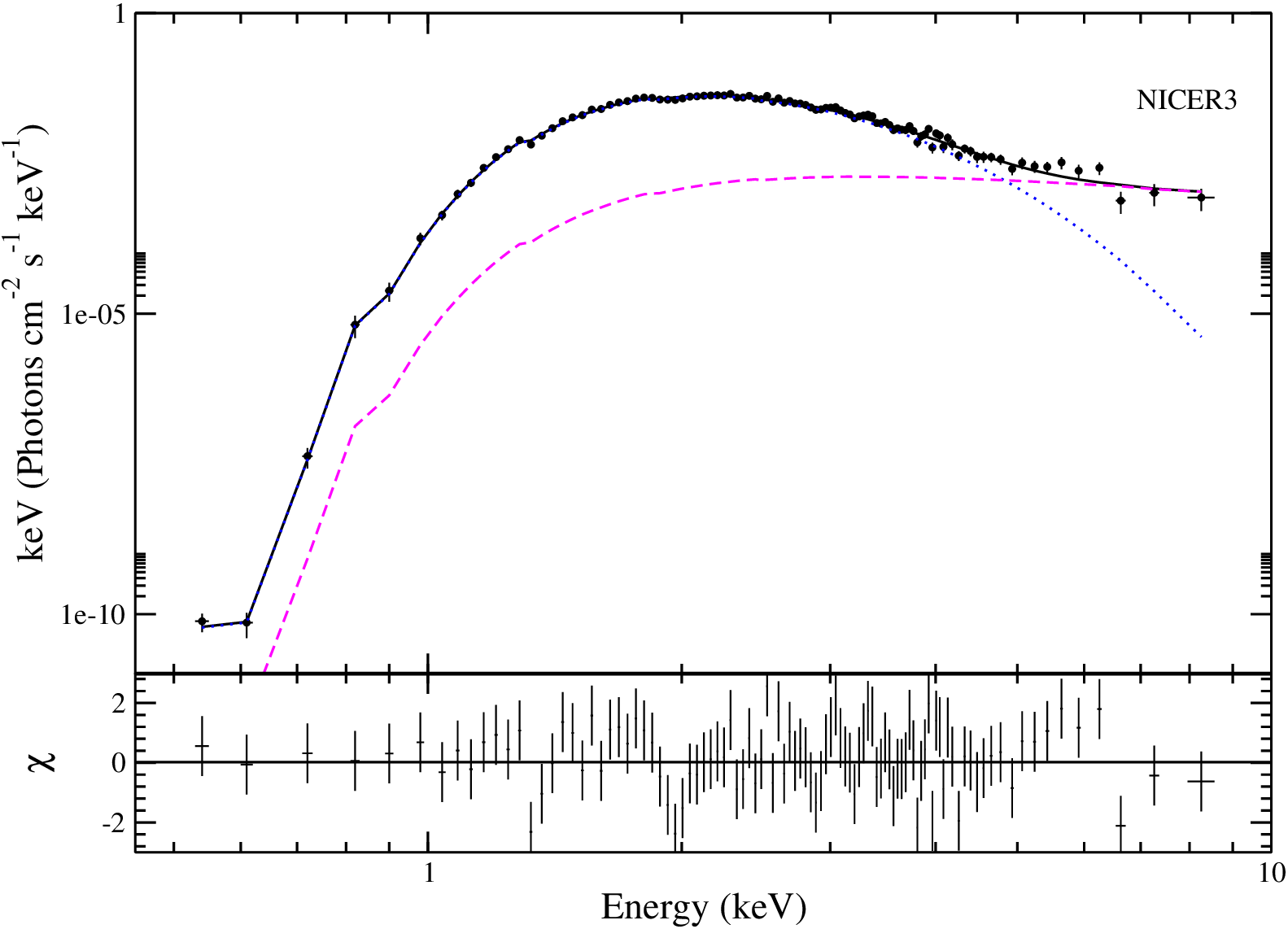}
    \includegraphics[width=7.5cm]{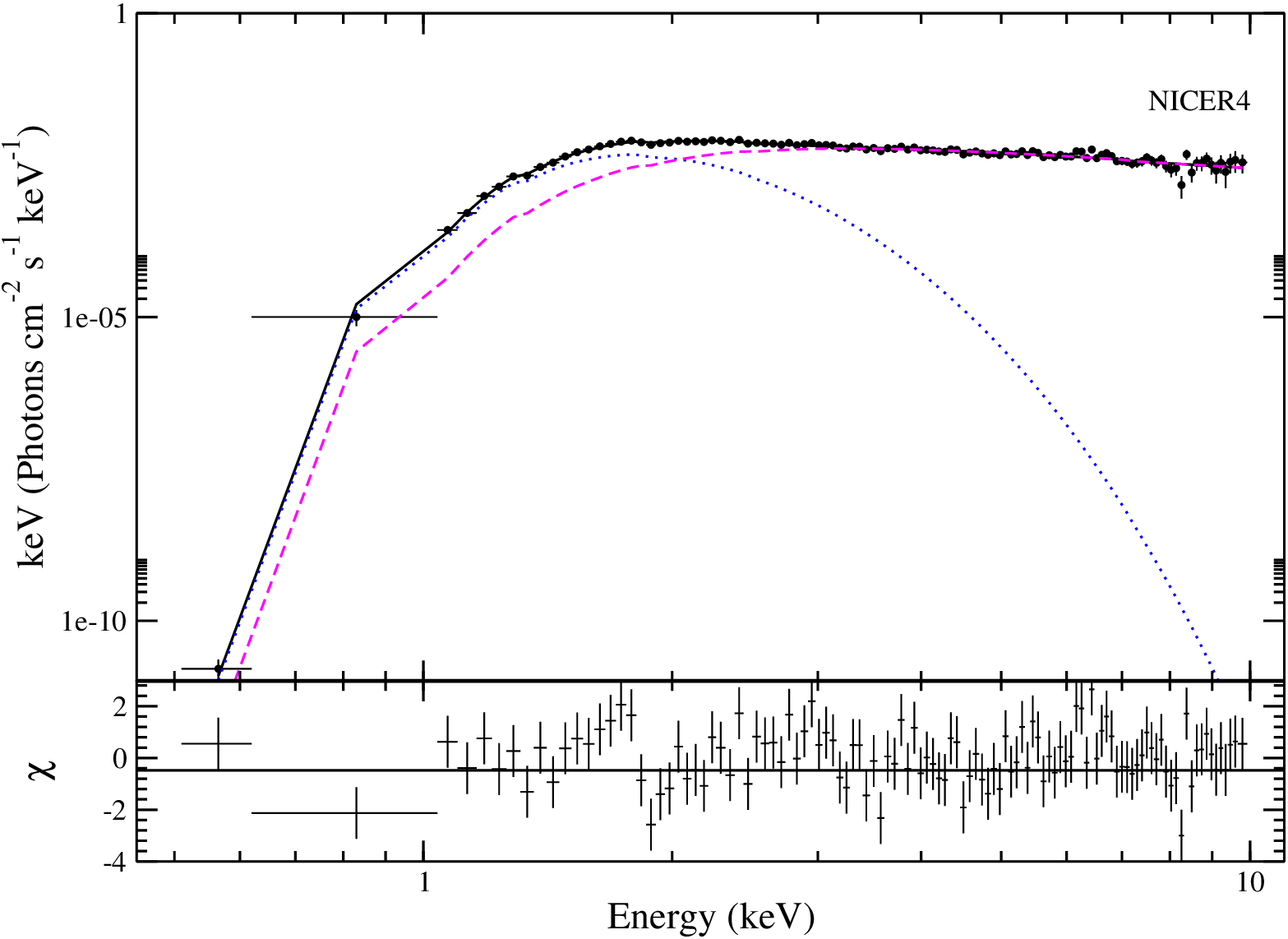}
    \caption{\textsc{tbabs*(DiskBB+PL)} model fitted NICER ($0.5-10$ keV) spectra with their corresponding residuals. The lower panels represent the $\chi^2$ variation for fitted models. The total model spectra are solid black lines in all panels. Dotted blue lines are the \textsc{DiskBB} components. The magenta dotted lines represent the \textsc{PL} components.}
    \label{fig:spec_nicer}
\end{figure}

\begin{figure}[!ht]
\centering
	\includegraphics[width=7.5cm]{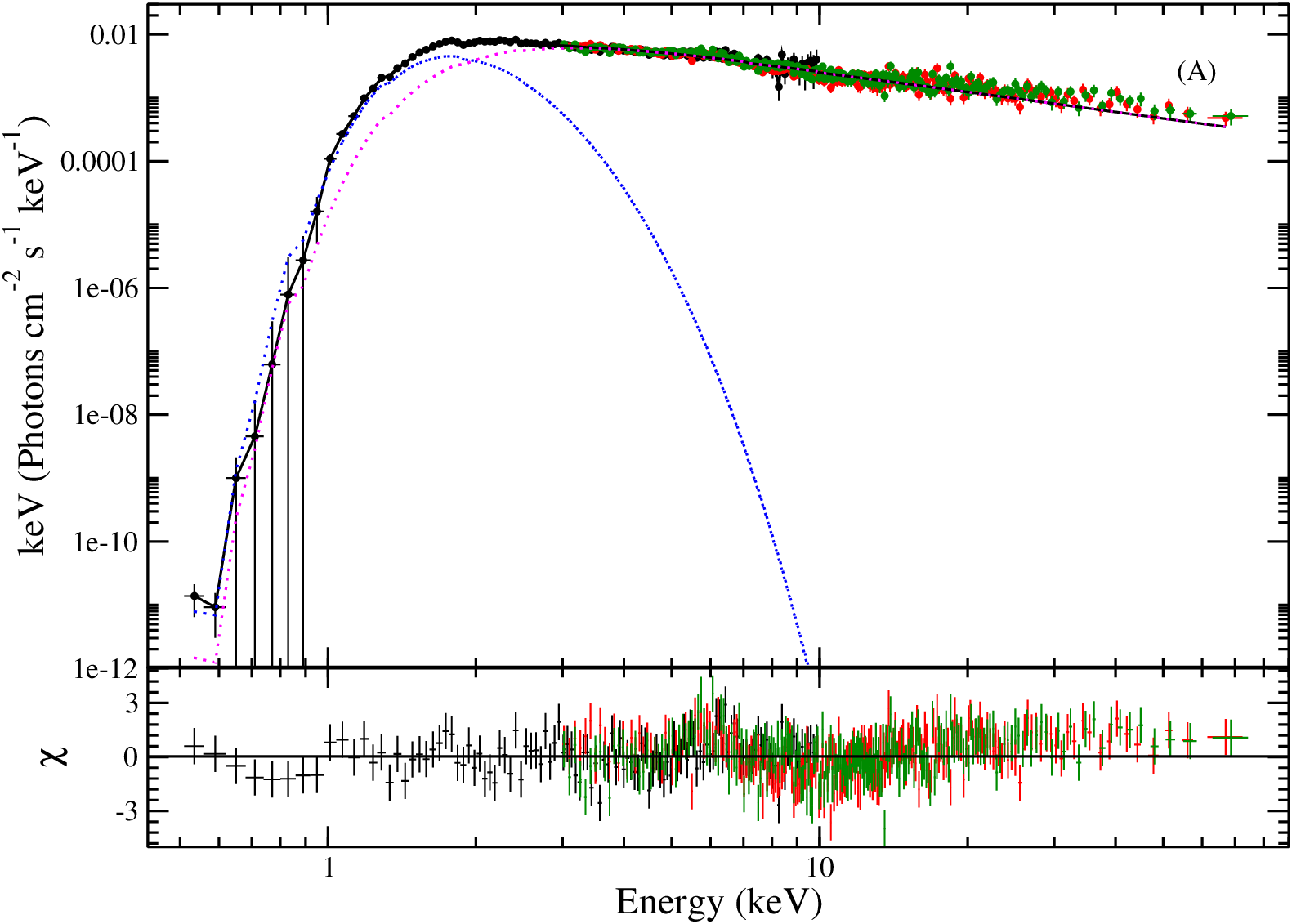}
 	\includegraphics[width=7.5cm]{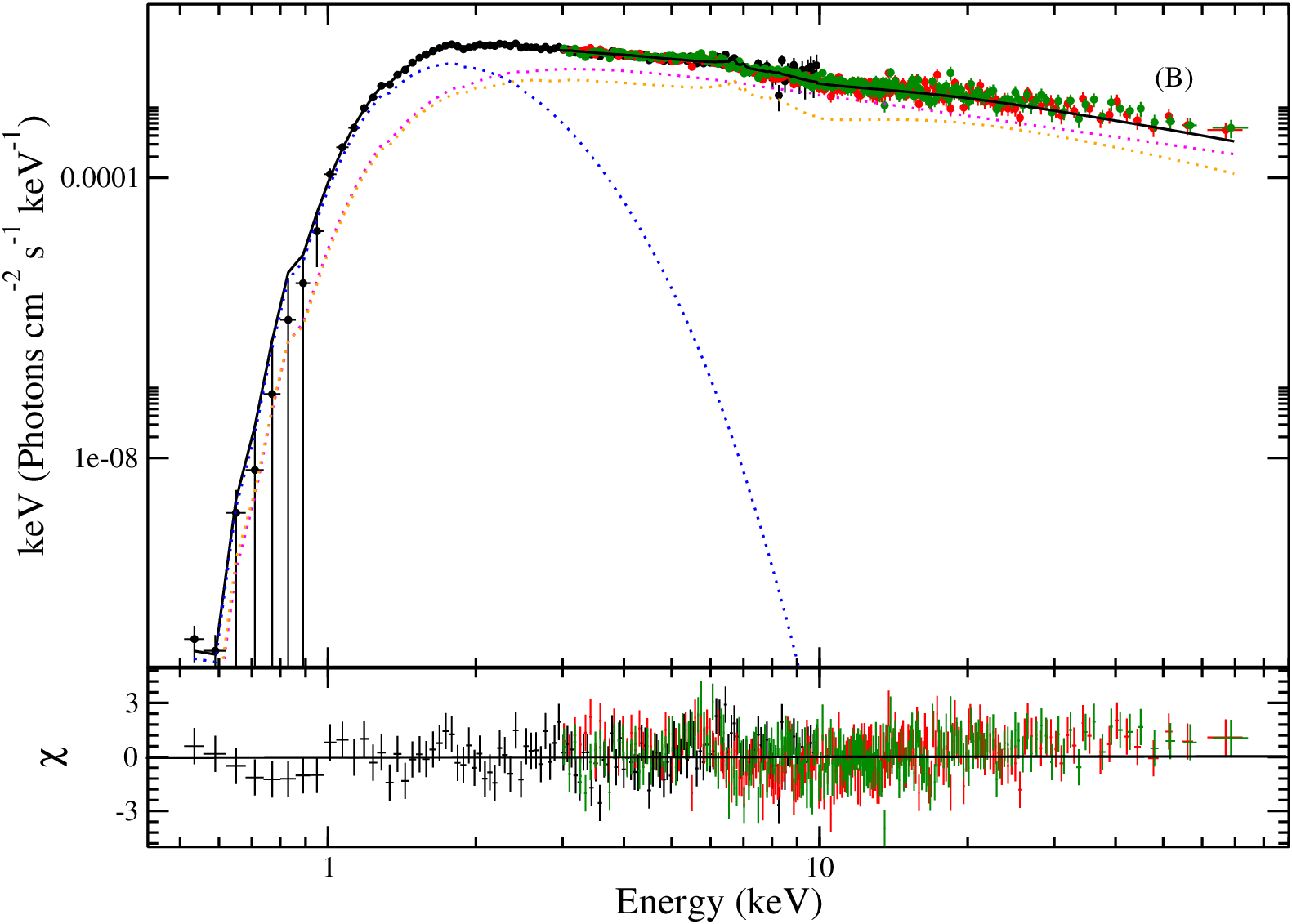}
    \includegraphics[width=7.5cm]{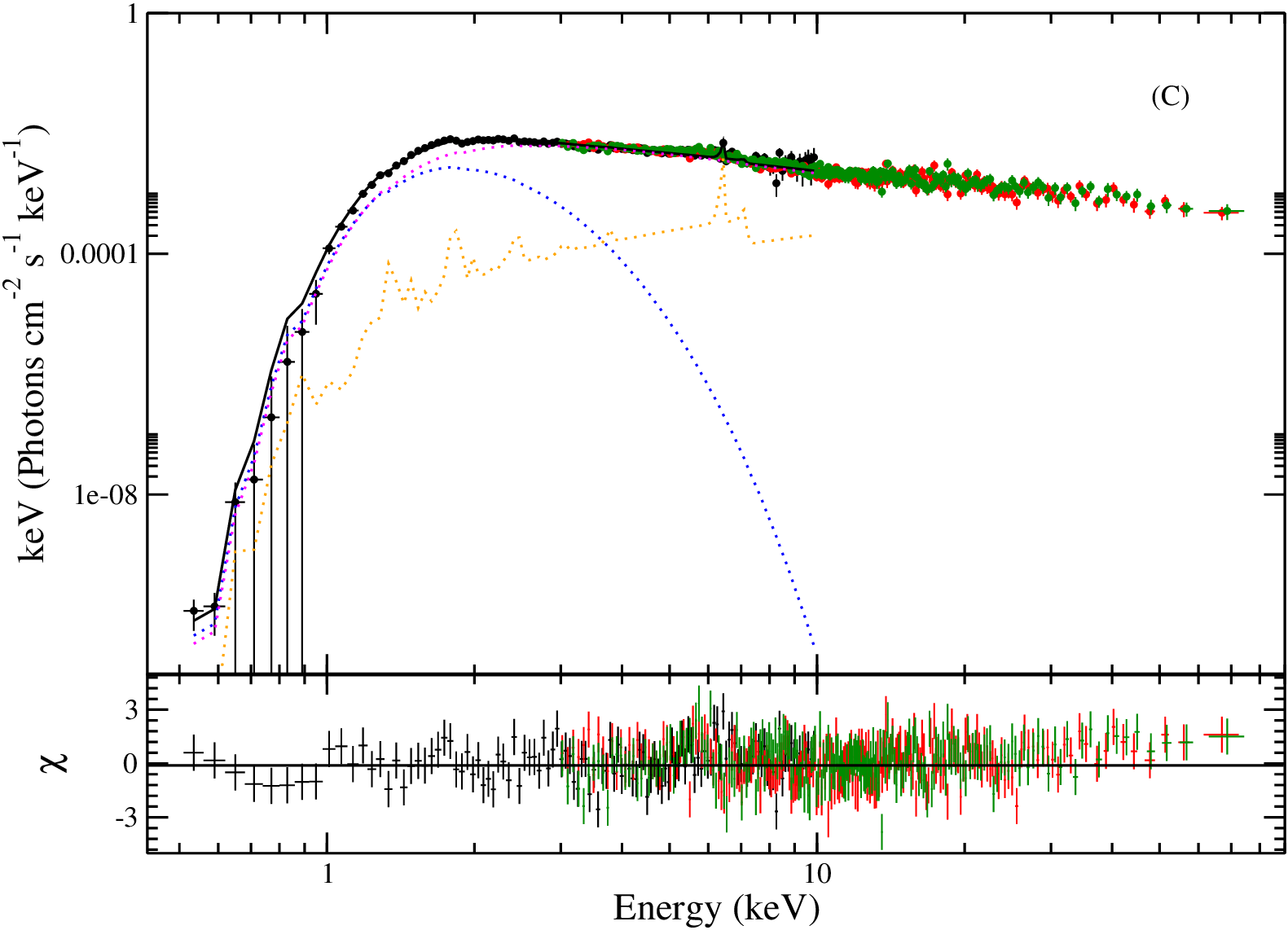}
    \includegraphics[width=7.5cm]{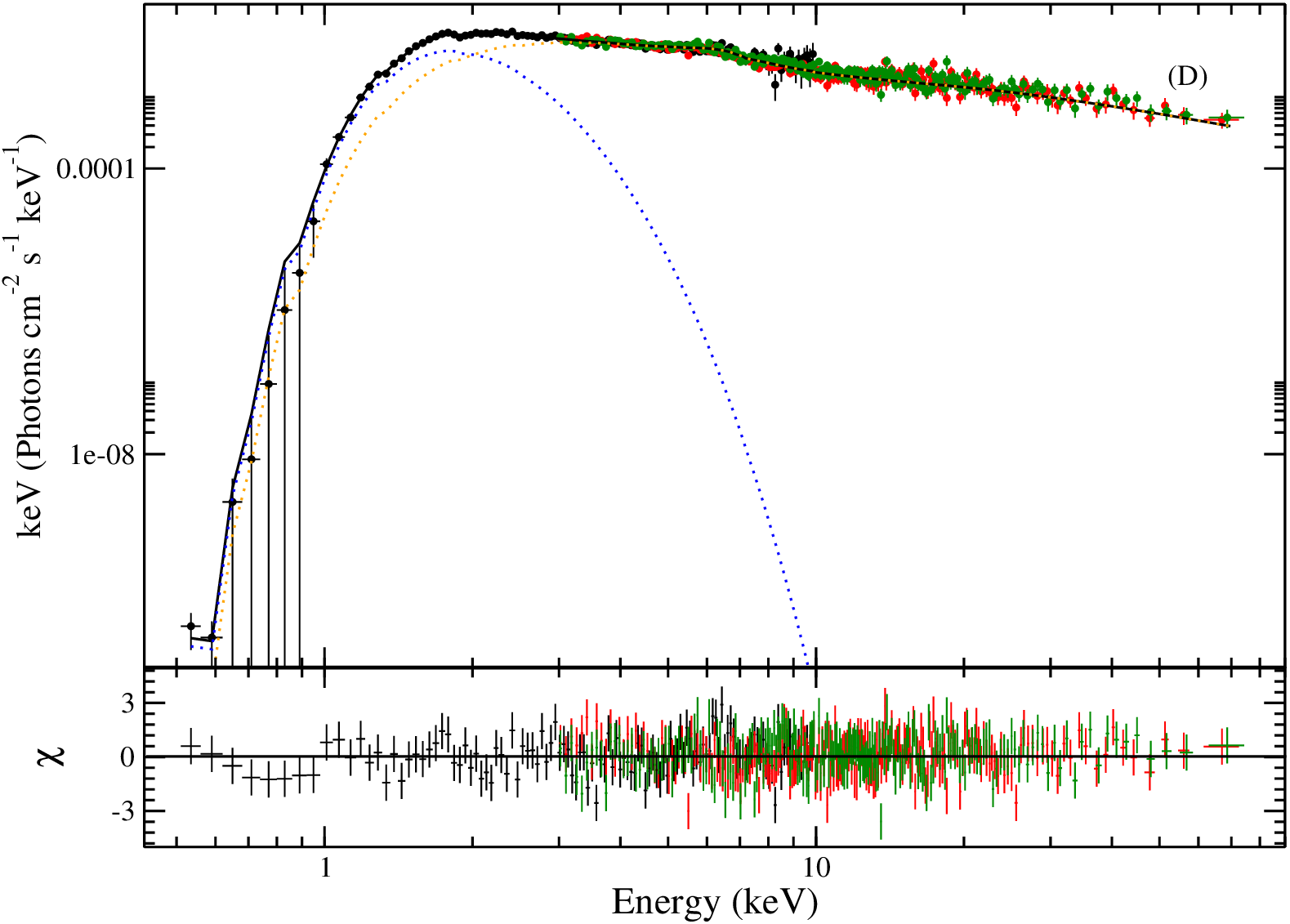}
    \caption{Model fitted combined NICER4 ($0.5-10$ keV) and NuSTAR ($3-78$ keV) spectrum with their corresponding residuals. Black, red and green represent the data of {\it NICER}, {\it NuSTAR} FPMA and FPMB, respectively. The lower panels represent the $\chi^2$ variation for fitted models. A, B, C and D corresponds to \textsc{tbabs*(DiskBB+PL)}, \textsc{Tbabs(DiskBB+PL+xillver)}, \textsc{Tbabs(DiskBB+ThComp~$\otimes$~DiskBB+xillver)}, and \textsc{tbabs*(DiskBB+relxill)} spectra respectively. The total model spectra are solid black lines in all panels (A, B, C, D). Dotted blue lines are the \textsc{DiskBB} components. \textsc{xillver} component is shown by orange dotted lines in panels B and C. The magenta dotted lines represent the \textsc{PL} components in panels A and B. In panel C, magenta \textsc{ThComp$\otimes$DiskBB} component. In panel D, the orange dotted line represents the \textsc{relxill} component.}
    \label{fig:comb_spectra_low_energy}
\end{figure}


\bibliography{sample631}{}
\bibliographystyle{aasjournal}



\end{document}